\documentclass[11pt]{article}
\usepackage[T1]{fontenc}
\usepackage[latin9]{inputenc}
\usepackage{geometry}
\usepackage{color}
\usepackage{float}
\usepackage{amsmath}
\usepackage{amsthm}
\usepackage{amssymb}
\usepackage{graphicx}
\usepackage{setspace}
\usepackage[authoryear]{natbib}
\usepackage{amsfonts}
\usepackage{dsfont}
\usepackage{mathtools}
\usepackage{booktabs}
\usepackage{multirow}
\DeclarePairedDelimiter\floor{\lfloor}{\rfloor}

\makeatletter
\newcommand{\id}{\mathds{1}}
\newcommand{\VaR}{\mathrm{VaR}}

\newcommand{\ES}{\mathrm{ES}}
\def\d{\mathrm{d}}
\providecommand{\tabularnewline}{\\}

\theoremstyle{definition}
    \ifx\thechapter\undefined
      \newtheorem{defn}{\protect\definitionname}
    \else
      \newtheorem{defn}{\protect\definitionname}[chapter]
    \fi
\theoremstyle{remark}
    \ifx\thechapter\undefined
      \newtheorem{rem}{\protect\remarkname}
    \else
      
    \fi
\theoremstyle{plain}
    \ifx\thechapter\undefined
	    \newtheorem{thm}{\protect\theoremname}
	  \else
      \newtheorem{thm}{\protect\theoremname}[chapter]
    \fi
\theoremstyle{plain}
    \ifx\thechapter\undefined
      \newtheorem{lem}{\protect\lemmaname}
    \else
      \newtheorem{lem}{\protect\lemmaname}[chapter]
    \fi
\theoremstyle{plain}
    \ifx\thechapter\undefined
      \newtheorem{prop}{\protect\propositionname}
    \else
      \newtheorem{prop}{\protect\propositionname}[chapter]
    \fi
\theoremstyle{definition}
    \ifx\thechapter\undefined
      \newtheorem{example}{\protect\examplename}
    \else
      \newtheorem{example}{\protect\examplename}[chapter]
    \fi
\theoremstyle{plain}
    \ifx\thechapter\undefined
  \newtheorem{cor}{\protect\corollaryname}
\else
      \newtheorem{cor}{\protect\corollaryname}[chapter]
    \fi

\usepackage{babel}

\@ifundefined{showcaptionsetup}{}{%
 \PassOptionsToPackage{caption=false}{subfig}}
\usepackage{subfig}
\makeatother

\usepackage{babel}
\providecommand{\corollaryname}{Corollary}
\providecommand{\definitionname}{Definition}
\providecommand{\examplename}{Example}
\providecommand{\lemmaname}{Lemma}
\providecommand{\propositionname}{Proposition}
\providecommand{\remarkname}{Remark}
\providecommand{\theoremname}{Theorem}
\newcommand{\X}{\mathcal{X}}
\newcommand{\Q}{\mathbb{Q}}
\newcommand{\A}{\boldsymbol{\alpha}}
\newcommand{\BB}{\boldsymbol{\beta}}
\newcommand{\W}{\boldsymbol{W}}
\newcommand{\E}{\mathbb{E}}
\newcommand{\Y}{\mathcal Y}
\newcommand{\R}{\mathbb R}
\newcommand{\w}{\mathbf w}
\newcommand{\B}{\mathbf B}

\newcommand{\thedate}{\today}
\usepackage{setspace}
\newcommand{\com}[1]{\marginpar{{\begin{minipage}{0.18\textwidth}{\setstretch{1.1} \begin{flushleft} \footnotesize \color{red}{#1} \end{flushleft} }\end{minipage}}}}

\usepackage{footmisc}
\providecommand{\tabularnewline}{\\}

\begin{document}
\global\long\def\com#1{\marginpar{\begin{minipage}{0.12\textwidth}{\setstretch{1.1} \begin{flushleft} \footnotesize{\color{red}{#1} }\end{flushleft} }\end{minipage}}}%

\title{Factor risk measures}
\author{
Hirbod Assa\thanks{School of Mathematics, Statistics and Actuarial Science, University of Essex, Colchester CO4 3SQ,  UK. Email: h.assa@essex.ac.uk}~ and
Peng Liu\thanks{School of Mathematics, Statistics and Actuarial Science, University of Essex, Colchester CO4 3SQ,  UK. Email: peng.liu@essex.ac.uk}}
 \date{\thedate}
\maketitle
\begin{abstract}
This paper introduces and studies factor risk measures. While risk
measures only rely on the distribution of a loss random variable, in many cases risk needs to
be measured relative to some major factors. In this paper, we introduce
a double-argument mapping as a risk measure to assess the risk relative to a vector of factors, called factor risk measure. The factor risk measure only depends on the joint distribution of the risk and the factors.  A set of natural
axioms are discussed, and particularly distortion, quantile,
 linear and coherent factor risk measures are introduced and characterized.
Moreover, we introduce a large set of concrete factor risk measures and many of them are new to the literature, which are interpreted in the context of regulatory capital requirement. Finally,  the distortion factor risk measures are applied in the risk-sharing problem and
 some numerical examples are presented to show
the difference between the Value-at-Risk and the quantile factor risk measures.
\end{abstract}
\begin{onehalfspace}
\end{onehalfspace}

\section{Introduction}

~~~~~Risk measures have been widely used in banking and insurance for various purposes such as regulation, optimization and risk pricing. Since the seminal paper of \cite{ADEH99}, risk measures are commonly defined as functionals on a set of random variables representing profit/loss of the portfolios. The commonly used risk measures are law-invariant, only depending on the distribution of the risk.  Two popular law-invariant risk measures in regulation are the Value-at-Risk ($\VaR$) and  the Expected Shortfall ($\ES$); one can refer to  \cite{MFE15} and \cite{FS16} for more discussions on the risk measures.

From the regulatory prospective, a regulator is more concerned with the risk in stress scenario; see e.g., \cite{AER12} for more detailed discussion.  In the credit rating practice, a structured finance security is rated by the risk behavior on each economic scenario; see e.g., \cite{SP19}, \cite{M23} and \cite{GKWW}. However, the distribution of the risk alone is not able to capture the behavior of the risk under different scenarios. Different scenarios can be represented by some random variables. Those random variables can be viewed as factors. The nature of the risk is more comprehensively described by the risk together with some factors than the risk alone. Hence,
different problems in insurance and finance consider risk management
relative to a set of specific risk factors. For instance, in the systematic risk, the market risk works as the factor;
in pricing, the economic factors indexes are considered as the factors; in the systemic risk, the
common shock is the variable relative to which we do the risk management; in the catastrophic risk, specific natural indexes such as the
temperature in frost events, or Richter index in earthquake constitute
the major factors for the losses.

All the above facts motivate us to study risk functionals with double arguments: one is the risk and the second is a vector of random variables representing the factors. We call them factor risk measures. Recently, in \cite{WZ21}, different scenarios are captured by a set of probability measures,
   which is very different but also closely related to the setup in this paper. One can also refer to \cite{KP16} for the similar consideration.  More details will be discussed in Sections \ref{sec:preliminary} and \ref{sec:distortion}.

Since the
financial crisis in 2008, systemic risk have been considered
more seriously as part of the risk management process. Co-risk measures are among the most studied systemic risk measures. This
type of risk measures conditions the risk measurement on the occurrence
of a systemic risk such as  conditional Value-at-Risk (CoVaR)
of \cite{AB16},
the conditional Expected Shortfall (CoES) of \cite{MS14},
and the marginal Expected Shortfall (MES) of \cite{APPR17}.
Recently, \cite{DLZ22} have integrated all the
above risk measures in one, and call them conditional distortion risk
measure.  The Co-risk measures actually rely on the conditional distribution of the risk on the event of systemic risk, which is determined by the joint distribution of the risk and the systemic risk. This is intrinsically different from the classical law-invariant risk measures in \cite{FS16}. In this paper, we will follow the same idea as Co-risk measures to consider the factor risk measures solely depending on the joint distribution of the risk and the factors. We call it law-invariant factor risk measure.

Compared with the risk measures in the literatures, the novelty of this paper stems from the new type of law-invariance.  The law-invariance considered in this paper is closely related to the scenario-based law-invariance discussed in \cite{WZ21} and the conditional law-invariance studied in \cite{DE21} and \cite{de24}. Their relation is discussed in Proposition \ref{law-invariance} in Section \ref{sec:preliminary}.
The law-invariance is crucial in our characterization as it relates the risk and the factors and allows us to evaluate the risk relative to the factors. Hence, the main theme of this study
is to assess the risk relative to a set of major factors. This
will be clear from our discussions and examples later in the paper.

 Under this new law-invariance, we study factor risk measures satisfying  monotonicity and comonotonic-additivity which are called the distortion factor risk measures in Section \ref{sec:distortion}. The classical distortion risk measures have been widely applied in  decision theory (\cite{Y87}), insurance and option pricing (\cite{Wang96} and \cite{Wang00}), performance evaluation (\cite{CM09}) and quantitative risk management (\cite{MFE15} and \cite{FS16}). Hence, its generalization to the case with factors is important from both theoretical and practical perspective. In our characterization in Theorem \ref{thm:choquet}, the distortion factor risk measure can be represented by a Choquet integral defined by a distortion functional on a set of Borel measurable functions, which extends the classical distortion risk measures given by a Choquet integral with a distortion function on the real line.  Moreover, we find a necessary and sufficient condition on the distortion functionals such that the distortion factor risk measure is coherent. Finally, we show some concrete examples of distortion factor risk measures such as $\mathrm{CoES}$, the distortion of conditional $\VaR$ and the expectation of conditional $\ES$. Most of those factor risk measures are new  and they offer new angles to evaluate the risk affected by some factors. To some extend, our characterization results in Section \ref{sec:distortion} extend the results in \cite{WZ21}. It is also worth mentioning that our characterization with the aid of the distortion functionals on a set of Borel measurable functions is very different from \cite{GHW22}, where the axioms are state-wise and the expressions are state-wise based.

Quantiles are one of the important tools in decision theory (\cite{R10} and \cite{CG19}), statistics (\cite{KH01}), finance (\cite{BS01}), risk management (\cite{ELW18}) and many other fields. Its characterization has been considered using different axioms such as elicitability in \cite{KP16}, tail-relevance in \cite{LW21}, and  ordinality in \cite{C09} and \cite{FLW22}.
 In Section \ref{sec:VaR},  we study factor risk measures satisfying law-invariance, monotonicity and ordinality, called quantile factor risk measures. We find the expressions of the quantile factor risk measures including $\mathrm{CoVaR}$ as a special case. The expressions in Theorem \ref{thm:MVaR} have the same spirit as the definition of quantiles, offering a natural extension of quantiles to the cases with factors. Some concrete examples are given such as the $\VaR$ of conditional $\VaR$ and the esssup of conditional $\VaR$. Those quantile factor risk measures are interpreted in the context of the regulatory capital requirement and they have the potential to be applied in practice because of the simple form.

  In Section \ref{Sec:linear}, we characterize linear factor risk measures, which can be expressed as the weighted average of the conditional expectation. It includes $\mathrm{MES}$ as a special case and has the same form as Moody's rating measure for credit rating (\cite{M23}). Coherent risk measures and its extensions become popular in the quantitative risk management since the seminal papers of \cite{ADEH99}, \cite{FS02} and \cite{FR05}; see also \cite{FS16} for a comprehensive overview of coherent risk measures.  The coherent factor risk measures are studied in Section \ref{sec:Coherent}.  It is nontrivial to obtain its expression as it relies on an extension of the Hardy-Littewood inequality. Some simple examples are also shown such as the $\ES$ of conditional $\ES$ and the esssup of conditional $\ES$.

Finally, in Section \ref{sec:risksharing}, we put our focus on the applications of the distortion factor risk measures to the comonotonic risk sharing problem and on the numerical comparison of $\VaR$ and the quantile factor risk measures in risk evaluation. Some notation and definitions are displayed in Section \ref{sec:preliminary} and all the proofs of the results in this paper are postponed to Appendix.

\section{Preliminaries}\label{sec:preliminary}

Consider an atomless probability space $\left(\Omega,\mathcal{F},\mathbb{P}\right)$,
where $\Omega$ is the set of the states, $\mathcal{F}$ is a sigma
field on $\Omega$ and $\mathbb{P}$ is a probability measure. Let
$X$ be a random variable, from the set $\mathcal{X}$ consisting
of loss variables. We suppose $\X=L^{p},~p\in[0,\infty]$. Let $\mathcal{Y}$
be the set of factor random variables consisting of random vectors
$\W=(W_{1},\dots,W_{N})$ with $N\geq1$. Typically, we suppose $\mathcal{Y}$
is law-invariant, i.e., $\W\in\Y$ implies $\W'\in\Y$ if $\W'\overset{d}{=}\W$,
where $\overset{d}{=}$ represents quality in distribution. 
As usual, $F_{X}$ denotes the cumulative distribution
function (CDF) of $X$ and $F_{X}^{-1}$ denotes the left quantile of $F_{X}$, given
by
\[
F_{X}^{-1}(\alpha)=\inf\{x\in\R:F_{X}(x)\geq\alpha\},~\alpha\in(0,1]
\]
with the convention $\inf\emptyset=\infty$. Moreover, two regulatory
risk measures Value-at-Risk ($\VaR$) and Expected shortfall ($\ES$)
are defined as follows: For $X\in L^{0}$,
\[
\mathrm{VaR}_{\alpha}\left(X\right)=F_{X}^{-1}\left(\alpha\right),~\alpha\in(0,1];
\]
For $X\in L^{1}$,
\[
\ES_{\alpha}(X)=\frac{1}{1-\alpha}\int_{\alpha}^{1}F_{X}^{-1}(t)\d t,~\alpha\in[0,1).
\]
 Another important class of risk measure is called distortion risk measure.
For $X\in L^\infty$, the distortion risk measure with the distortion function $\Lambda\in \mathfrak{D}$ is given by
$$\varrho_{\Lambda}(X)=\int_{0}^{\infty} \Lambda(1-F_X(x))\d x+\int_{-\infty}^{0}\left(\Lambda(1-F_X(x))-1\right)\d x,$$
where  $\mathfrak{D}$ denotes the set of all non-decreasing functions $\Lambda:[0,1]\to[0,1]$ satisfying $\Lambda(0)=0$ and $\Lambda(1)=1$.
We refer to  \cite{MFE15} and \cite{FS16} for more discussions on these
 risk measures.

Before we introduce the conditional $\VaR$ and $\ES$, we need the following nation. For $X$ and $\W$, by Theorem 33.3 of \cite{B95}, there exists $K:\mathcal B(\R)\times \mathcal B(\R^N)$ such that
\begin{enumerate}
\item[(i)] For each $\w\in \R^N$, $K(\cdot, \w)$ is a probability measure on $\mathcal B(\R)$;
\item[(ii)] For each $B\in \mathcal B(\R)$, $K(B,\W)$ is a version of $\mathbb P(X\in B|\W)$.
\end{enumerate}
The conditional quantile is defined by $\VaR_{\alpha}(X|\W)=\inf\{x\in\R: K((-\infty,x],\W)\geq \alpha\}$ for $\alpha\in (0,1]$ and $\VaR_{0}(X|\W)=\inf\{x\in\R: K((-\infty,x],\W)>0\}$. Moreover, the conditional expected shortfall (ES) is defined by for $X\in L^1$, $\ES_\alpha(X|\W)=\frac{1}{1-\alpha}\int_{\alpha}^1 \VaR_{t}(X|\W)\d t$ for $\alpha\in [0,1)$.

In \cite{AB16} and \cite{MS14}, the $\mathrm{CoVaR}$ and $\mathrm{CoES}$ are defined as below:
For $\alpha,\beta\in(0,1)$ and $X,W\in\mathcal{X}$, $$\mathrm{CoVaR}_{\alpha,\beta}(X,W)=\mathrm{VaR}_{\beta}(X|W\geq\mathrm{VaR}_{\alpha}(W)),$$
$$\mathrm{CoVaR}_{\alpha,\beta}^{=}(X,W)=\mathrm{VaR}_{\beta}(X|W=\mathrm{VaR}_{\alpha}(W)),$$  and
$$\mathrm{CoES}_{\alpha,\beta}(X,W)=\frac{1}{1-\beta}\int_{\beta}^{1}\mathrm{CoVaR}_{\alpha, t}(X|W)\d t,$$
$$\mathrm{CoES}_{\alpha,\beta}^{=}(X,W)=\frac{1}{1-\beta}\int_{\beta}^{1}\mathrm{CoVaR}_{\alpha, t}^{=}(X|W)\d t.$$
For $\alpha\in (0,1)$, the MES is given by  $\mathrm{MES}_{\alpha}(X,W)=\mathbb E(X|W\geq \mathrm{VaR}_{\alpha}(W))$ in \cite{APPR17}.
Our characterization of factor risk measures in Sections \ref{sec:distortion}-\ref{sec:linear} cover those Co-risk measures as special cases.

We next  introduce the factor risk measure. Our approach following the same idea as in the literature is to introduce axioms that are
perceived to be the most appealing ones to insurance and finance applications,
and then explore the implications. The axioms that are being presented
are motivated from three major strands, a group stemming from the distortion
risk measures, a group from the quantiles,  and a group stemming from coherent risk measures.

Let us consider a risk $X$ whose risk measurement is our
main objective and a vector of factors , $\W=\left(W_{1},...,W_{N}\right)$,
that has a great influence in the risk measurement of variable $X$.
These factors can have different interpretations, for instance, they
can be macro-economic factors, they can be the systemic or systematic
risk variable or just some random factors that may add to the uncertainty
of the model.
\begin{defn}
A factor risk measure is a double-argument functional $\rho:\mathcal{X}\times\mathcal{Y}\to\mathbb{R}$
to measure the risk of $X$ given $\W$, denoted by $\rho\left(X,\W\right)$.
The first variable $X$ is called the risk variable and the second
variable $\W$ is called the factor variable.
\end{defn}


We next introduce the axioms that are not only appealing in the application of insurance and finance but also  play crucial role in our characterization results later. We say $X_1,  X_2$ are \emph{comonotonic} if there exist a random variable $Z$ and two non-decreasing functions $f_1$ and $f_2$    satisfying $f_1(x)+f_2(x)=x,~x\in\R$ such that $X_1=f_1(Z)$ and $X_2=f_2(Z)$.
\begin{enumerate}
\item \textbf{Monotonicity (M).} For any two random variables $X\le Y$,
and any factor $\W$, we have $\rho\left(X,\W\right)\le\rho\left(Y,\W\right)$.
\item \textbf{Comonotonic additivity (CA).} For any factor $\W$ and two
comonotone random variables $X_1$ and $X_2$, we have $\rho\left(X_{1}+X_{2},\W\right)=\rho\left(X_{1},\W\right)+\rho\left(X_{2},\W\right).$
\item \textbf{Normalization (N).} For all $\W,\rho\left(1,\W\right)=1$.
\item \textbf{Law-invariance (LI).} If $(X,\W)\overset{d}{=}(X',\W')$, then $\rho\left(X,\W\right)=\rho\left(X',\W'\right)$.
\item \textbf{Ordinality (OR).} For all continuous and strictly increasing
functions $\phi:\R\to\R$, and $X\in\mathcal{X}, \W\in\mathcal Y$, we have
\[
\rho\left(\phi(X),\W\right)=\phi\left(\rho\left(X,\W\right)\right).
\]
\end{enumerate}
As one can realize axioms M, CA and N, are mainly stemmed from the
theory of distortion risk measures as it is shown in \cite{Sch86}. Axiom LI is very different from the corresponding concept in the literature, showing that the factor risk measure solely depends on the joint distribution of the risk and the factors.  OR is a property introduced in \cite{C07, C09} to the axiomatize quantiles. In the decision-theoretic setup, it means that any continuous and strictly increasing transform on the prospects does not change the preference order; see e.g., \cite{FLW22}.

We next introduce the \emph{coherent} factor risk measures.
For a mapping $\rho:\mathcal{X}\times\mathcal{Y}\to\mathbb{R}$, we say $\rho$ is \emph{cash-invariant} if $\rho(X+c,\W)=\rho(X,\W)+c$ for all $X\in\X$ and $c\in\R$; $\rho$ is \emph{positively homogeneous} if $\rho(\lambda X,\W)=\lambda\rho(X,\W)$ for all $\lambda\geq 0$ and $X\in\X$; $\rho$ is \emph{subadditive} if $\rho(X+Y,\W)\leq \rho(X,\W)+\rho(Y,\W)$ for $X,Y\in\X$.  We say $\rho$ is \emph{monetary} if $\rho$ is monotone and cash-invariant; $\rho$ is \emph{coherent} if $\rho$ is monetary, positively homogeneous and subadditive.  We refers to the monograph \cite{FS16} for more details on the coherent risk measures.

Here, it is worth emphasizing that the novelty of this paper comes from LI.
The law-invariance discussed
in this paper is closely related to the $\mathcal Q$-based law-invariance introduced in \cite{WZ21}.
For a collection of probability measures $\mathcal Q$, we say $\varrho:\X \to\R$ is $\mathcal{Q}$-based if for
$X, Y\in \X$, $X
\overset{d}{=}
Y$ under $Q$ for all $Q\in \mathcal Q$ implies $\varrho(X)=\varrho(Y)$.
The mapping $\rho$ considered in this paper is a double-argument mapping. However, if we fix the second argument, it boils down to a single-argument mapping. We obtain the following conclusion on the law-invariance defined in this paper and the $\mathcal Q$-based law-invariance in \cite{WZ21}.
\begin{prop}\label{law-invariance} Fix $\W\in\mathcal Y$.  A mapping $\rho(\cdot,\W) :\X \to\R$ is law-invariant if and only if it is $\mathcal Q_{\W}$-based,
where $\mathcal Q_{\W}=\left\{\mathbb P\left(\cdot\Big| \W\in \prod_{i=1}^N (a_i,b_i)\right): \mathbb P\left(\W\in \prod_{i=1}^N (a_i,b_i)\right)>0,~a_i,b_i\in\mathbb Q\right\}$; If $\W$ is discrete, then $\mathcal Q_{\W}$  can be chosen as $\mathcal Q_{\W}=\left\{\mathbb P\left(\cdot\big| \W=\w\right): \mathbb P\left(\W=\w\right)>0, \w\in\R^{N}\right\}$.
\end{prop}

In \cite{DE21} and \cite{de24}, the mapping $\varrho:\X\to\mathcal X'$ is studied based on
conditional law-invariance, where $\X'$ is a set of random variable.  The conditional law-invariance has the similar meaning to  the LI in this paper if we fix the second argument. 

In the rest of the paper, for simplicity and consistency, we set $\X=L^\infty$. Some of the results can be easily extended to more general sets.

\section{Distortion factor risk measures}
\label{sec:distortion}

In this section, we shall study the risk measures satisfying M and CA.  Let us first introduce some  notation which plays a crucial role in the characterization results in this paper.

Let $D^{*}$ be the set of all Borel measurable functions $f:\R^{N}\to[0,1]$,
and $G_{\W}:D^{*}\to[0,1]$ is a functional satisfying $G_{\W}(0)=0$
and $G_{\W}(1)=1$ for $\W\in\mathcal{Y}$. We say $G_{\W}$ is monotone
if for $f,g\in D^{*}$, $f\geq_{\mathbb{P}\circ\W^{-1}}g$, i.e.,
$\mathbb{P}(f(\W)\geq g(\W))=1$, implies $G_{\W}(f)\geq G_{\W}(g)$;
$\{G_{\W}:\W\in\mathcal{Y}\}$ is \emph{law-invariant} if $G_{\W}=G_{\W'}$ whenever $\W\overset{d}{=}\W'$ for $\W,\W'\in\mathcal{Y}$.
Note that for $A\in\mathcal{F}$ and $\W\in\mathcal{Y}$, $\mathbb{P}(A|\W=\cdot)$
is a Borel measurable function in $D^{*}$ and $\mathbb{P}(A|\W=\cdot)$
is not unique. We denote $\mathbb{P}(X\leq x|\W=\cdot)$ by $F_{X|\W=\cdot}(x)$.
\begin{thm}
\label{thm:choquet} A mapping $\rho:\mathcal{X}\times\mathcal{Y}\to\mathbb{R}$
satisfies conditions M, CA, N and LI if and only if there exists a law-invariant family of monotone functionals $\{G_{\W}:\W\in\mathcal{Y}\}$
such that
\begin{equation}
\rho(X,\W)=\int_{0}^{\infty}G_{\W}\left(1-F_{X|\W=\cdot}(x)\right)\d x+\int_{-\infty}^{0}\left(G_{\W}\left(1-F_{X|\W=\cdot}(x)\right)-1\right)\d x.\label{eq:quantiles}
\end{equation}
\end{thm}

It is noteworthy that $G_{\W}$ provides a natural
extension of distortion functions to a conditional one, and therefore
can be used to introduce the factor risk measures interpreted as
the risk measure based on the conditional distribution functions.

For a
family of monotone functionals $\{G_{\W}:\W\in\mathcal{Y}\}$, we denote \eqref{eq:quantiles}
by $\rho_{G_{\W}}(X,\W)$ and call it the \emph{distortion factor risk measure}
with distortion functionals $G_{\W}$ and factor $\W$. Clearly, $\rho_{G_{\W}}(X,\W)$ satisfies M, CA and N, which can be seen from the proof of Theorem \ref{thm:choquet} in Appendix \ref{Appendix:dis}.  Note that the law-invariance of $\{G_{\W}:\W\in\mathcal{Y}\}$ is a sufficient condition but not a necessary condition to ensure the law-invariance of $\rho_{G_{\W}}(X,\W)$.   A sufficient and necessary condition is  \emph{weak law-invariance} of $\{G_{\W}: \W\in\mathcal Y\}$: For $A, A'\in\mathcal F,\W, \W'\in\mathcal Y$ satisfying $(\id_A,\W)\overset{d}{=}(\id_{A'},\W')$, we have $G_{\W}(\mathbb P(A|\W=\cdot))=G_{\W'}(\mathbb P(A|\W=\cdot))$.
\begin{prop}\label{weakly law-invariant} For a distortion factor risk measure $\rho_{G_{\W}}(X,\W)$ with a family of monotone functionals $\{G_{\W}:\W\in\mathcal{Y}\}$, it is law-invariant if and only if $\{G_{\W}:\W\in\mathcal{Y}\}$ is weakly law-invariant.
\end{prop}

Note that the law-invariance of $\{G_{\W}:\W\in\mathcal{Y}\}$ implies the weak law-invariance of $\{G_{\W}:\W\in\mathcal{Y}\}$. However, the converse conclusion is not true in general. One can see a counterexample in Example \ref{example} in Appendix \ref{cexample}.

We say that $\rho$ is \emph{indicator lower semicontinuous}  (IC) if for $A_n, A\in\mathcal F$, $A_{n}\uparrow A,n\to\infty$,
implies that for all $\W\in\Y$,
\[
\rho(\id_{A},\W)\leq\liminf_{n\to\infty}\rho(\id_{A_{n}},\W).
\]

The following result shows that if $\rho$ additionally satisfies IC, then the corresponding $G_{\W}$ is \emph{continuous from below}: For $f_n\in D^*$, $f_n\uparrow f$ (pointwise) implies $\lim_{n\to\infty}G_{\W}(f_n)=G_{\W}(f)$. This result will be used to characterize the linear factor risk measure later in Section \ref{Sec:linear}.
\begin{prop}\label{prop:choquet} A mapping $\rho:\mathcal{X}\times\mathcal{Y}\to\mathbb{R}$
satisfies conditions  M, CA, IC, N and LI  if and only if there exists a law-invariant family of monotone and continuous from below functionals $\{G_{\W}: \W\in\mathcal Y\}$  such that \eqref{eq:quantiles} holds.
\end{prop}

If $\mathcal Y$ only contains random variables taking finite values, then Theorem \ref{thm:choquet} has a simpler version.
We say $\psi_{\W}:[0,1]^n\to [0,1]$ is an increasing function if for $\mathbf x, \mathbf y\in [0,1]^n$, $\psi_{\W}(\mathbf x)\leq \psi_{\W}(\mathbf y)$ whenever $\mathbf x\leq \mathbf y$, i.e., $x_i\leq y_i$ for all $i=1,\dots, n$; we say $\{\psi_{\W}:\W\in\mathcal{Y}\}$ is \emph{law-invariant} if $\psi_{\W}=\psi_{\W'}$ whenever $\W\overset{d}{=}\W'$ for $\W,\W'\in\mathcal{Y}$.
\begin{cor}\label{prop:discrete} Suppose all $\W\in \mathcal Y$ take $n$ different values.  A mapping $\rho:\mathcal{X}\times\mathcal{Y}\to\mathbb{R}$
satisfies conditions M, CA, N and LI if and only if there exists a law-invariant family of monotone functions $\{\psi_{\W}:\W\in\mathcal{Y}\}$
such that
\begin{align*}
\rho(X,\W)&=\int_{0}^{\infty}\psi_{\W}\left(1-F_{X|\W=\w_1}(x),\dots, 1-F_{X|\W=\w_n}(x)\right)\d x\nonumber\\
&+\int_{-\infty}^{0}\left(\psi_{\W}\left(1-F_{X|\W=\w_1}(x),\dots, 1-F_{X|\W=\w_n}(x)\right)-1\right)\d x.
\end{align*}
\end{cor}
 Note that the result in Corollary \ref{prop:discrete} corresponds to  Theorem 3.4 of \cite{WZ21} with mutually singular probability measures. More precisely, Theorem 3.4 of \cite{WZ21} states that for $\mathcal Q=\{P_1,\dots,P_n\}$ and $\ensuremath{\underbar{P}}=(P_1,\dots, P_n)$,  a mapping $\rho:\X\to\R$ is monetary,  commonotonic additive and $\mathcal Q$-based if and only if
  there is a function
$\psi:[0,1]^{n}\to[0,1]$ such that $\psi\circ\text{\ensuremath{\underbar{P}}}$
is standard and
\[
\rho(X)=\int_{0}^{\infty}\psi\circ\text{\ensuremath{\underbar{P}}}\left(1-F_{X}(x)\right)\d x+\int_{-\infty}^{0}\left(\psi\circ\text{\ensuremath{\underbar{P}}}\left(1-F_{X}(x)\right)-1\right)\d x.
\]
Here the mapping $\psi\circ\ensuremath{\underbar{P}}: \mathcal F\to [0,1]$ is standard means that it is increasing with set inclusion and $\psi\circ\text{\ensuremath{\underbar{P}}}\left(\emptyset\right)=1-\psi\circ\text{\ensuremath{\underbar{P}}}\left(\Omega\right)=0$.

For a fixed $\W\in\mathcal Y$, one can see from Proposition \ref{law-invariance} that in Theorem \ref{thm:choquet}, we are in fact dealing with $\mathcal Q_{\W}$-based risk measures, where $\mathcal Q_{\W}$ contains infinite many distinct probability measures. From this perspective, our result in Theorem \ref{thm:choquet}  extends Theorem 3.4 of \cite{WZ21}. The main technical challenge
of the extension of the statement in \cite{WZ21} is that one cannot easily define a function $\psi$ on $[0,1]^{\mathbb{N}}$ similarly as $G_{\W}$ in our paper. Note also that Corollary \ref{prop:discrete} has the same form as the rating measure of credit rating criteria in Theorem 7 of \cite{GKWW}.

One natural question is what conditions are needed to guarantee that
$\rho_{G_{\W}}(X,\W)$ is coherent.
Inspired by
Proposition 3.5 in \cite{WZ21} and Theorem 3.12.2 of \cite{MS02}, we introduce the following
condition. Condition A: For $f_{1},f_{2},g_{1},g_{2}\in D^{*}$ satisfying
$g_{1}\leq f_{1},f_{2}\leq g_{2}$ and $f_{1}+f_{2}=g_{1}+g_{2}$,
we have $G_{\W}(f_{1})+G_{\W}(f_{2})\geq G_{\W}(g_{1})+G_{\W}(g_{2})$
for all $\W\in\mathcal{Y}$.
\begin{prop}
\label{prop:coherent} Suppose $\rho:\mathcal{X}\times\mathcal{Y}\to\mathbb{R}$
is given by \eqref{eq:quantiles} with a law-invariant family of monotone functionals $\{G_{\W}:\W\in\mathcal{Y}\}$. Then $\rho$ is a coherent factor risk measure
if and only if $G_{\W}$ satisfies condition A.
\end{prop}
Distortion factor risk measure is very fruitful as it contains many interesting factor risk measures. We next display some examples with conditional $\VaR$ and $\ES$ as the building blocks.
\begin{example}\label{exam:distortion}
\begin{itemize}
\item[(i)] For $g\in D^*$ with $\mathbb P(0<g(\W)<1)=1$, if $G_{\W}(f)=\Lambda(\mathbb P(f(\W)>1-g(\W))$ for all $f\in D^*$, then
\begin{align*}\rho_{G_{\W}}(X,\W)&=\int_{0}^{\infty}\Lambda(\mathbb P(F_{X|\W}(x)<g(\W)))\d x+\int_{-\infty}^{0}\left(\Lambda(\mathbb P(F_{X|\W}(x)<g(\W)))-1\right)\d x\\
&=\varrho_{\Lambda}(\VaR_{g(\W)}(X|\W)).
\end{align*}
\item[(ii)] For $p\in (0,1)$, if $G_{\W}(f)=\Lambda(\mathbb P(f(\W)>1-p))$ for all $f\in D^*$, then
\begin{align*}\rho_{G_{\W}}(X,\W)=\varrho_{\Lambda}(\VaR_{p}(X|\W)).
\end{align*}
\item[(iii)] For $g\in D^*$ with $\mathbb P(0<g(\W)<1)=1$, if $G_{\W}(f)=\mathbb P(f(\W)>1-g(\W))$ for all $f\in D^*$, then
\begin{align*}\rho_{G_{\W}}(X,\W)&=\int_{0}^{\infty}\mathbb E(\id_{\{F_{X|\W}(x)<g(\W)\}})\d x+\int_{-\infty}^{0}\left(\mathbb E(\id_{\{F_{X|\W}(x)<g(\W)\}})-1\right)\d x\\
&=\mathbb E(\VaR_{g(\W)}(X|\W)).
\end{align*}
\item[(iv)] For $p\in (0,1)$, if $G_{\W}(f)=\mathbb P(f(\W)>1-p)$ for all $f\in D^*$, then
\begin{align*}\rho_{G_{\W}}(X,\W)=\mathbb E(\VaR_{p}(X|\W)).
\end{align*}
\end{itemize}
\end{example}

 For (i)-(iv), they can be interpreted as follows. For each scenario $\{\W=\w\}$, the capital requirement is computed using $\VaR$ at level $g(\w)$, where the level varies as the scenario changes. Then, $\mathbb E(\VaR_{g(\W)}(X|\W))$ represents the average capital requirement across all scenarios and $\varrho_{\Lambda}(\VaR_{g(\W)}(X|\W))$ represents the distorted average capital requirement across all scenarios.


Next, we see some examples related to conditional $\ES$.
\begin{example}\label{example:coherent}
\begin{itemize}
\item[(i)] For $g\in D^*$ with $\mathbb P(g(\W)<1)=1$, let  $G_{\W}(f)=\mathbb E\left(\frac{f(\W)\wedge (1-g(\W))}{1-g(\W)}\right)$ for all $f\in D^*$, then
    \begin{align*}\rho_{G_{\W}}(X,\W)&=\int_{0}^{\infty}\mathbb E\left(\frac{(1-F_{X|\W}(x))\wedge (1-g(\W))}{1-g(\W)}\right)\d x\\
    &+\int_{-\infty}^{0}\left(\mathbb E\left(\frac{(1-F_{X|\W}(x))\wedge (1-g(\W))}{1-g(\W)}\right)-1\right)\d x\\
    &=\mathbb E\left(\ES_{g(\W)}(X|\W)\right).
    \end{align*}
\item[(ii)] For $p\in [0,1)$,  let  $G_{\W}(f)=\mathbb E\left(\frac{f(\W)\wedge (1-p)}{1-p}\right)$ for all $f\in D^*$, then
    \begin{align*}\rho_{G_{\W}}(X,\W)=\mathbb E\left(\ES_{p}(X|\W)\right).
    \end{align*}
\item[(iii)]  For $p\in [0,1)$, and $\mathbf B\in\mathcal{B}(\R^N)$ with $\mathbb P(\W\in\mathbf B)>0$, where $\mathcal{B}(\R^{N})$ represents the collection of all Borel
subsets of $\R^{N}$, let  $G_{\W}(f)=\mathbb E\left(\frac{f(\W)\wedge (1-p)}{1-p}\Big |\W\in\mathbf B\right)$ for all $f\in D^*$, then
    \begin{align*}\rho_{G_{\W}}(X,\W)&=\int_{0}^{\infty}\mathbb E\left(\frac{(1-F_{X|\W}(x))\wedge (1-p)}{1-p}\Big |\W\in\mathbf B\right)\d x\\
    &+\int_{-\infty}^{0}\left(\mathbb E\left(\frac{(1-F_{X|\W}(x))\wedge (1-p)}{1-p}\Big |\W\in\mathbf B\right)-1\right)\d x\\
    &=\ES_{p}(X|\W\in\mathbf B),
    \end{align*}
    which extends Co$\ES$.
\item[(iv)]  For $g\in D^*$ with $\mathbb P(g(\W)<1)=1$, let  $G_{\W}(f)=\mathrm{ess}\sup\left(\frac{f(\W)\wedge (1-p)}{1-p}\right)$ for all $f\in D^*$, then
    \begin{align*}\rho_{G_{\W}}(X,\W)&=\int_{0}^{\infty}\mathrm{ess}\sup\left(\frac{(1-F_{X|\W}(x))\wedge (1-p)}{1-p}\right)\d x\\
    &+\int_{-\infty}^{0}\left(\mathrm{ess}\sup\left(\frac{(1-F_{X|\W}(x))\wedge (1-p)}{1-p}\right)-1\right)\d x\\
    &=\ES_{p}(Y),
    \end{align*}
    where $\mathbb P(Y\leq x)=\lim_{y\downarrow x}\mathrm{ess}\inf F_{X|\W}(y)$ for $x\in\R$.  Clearly, $\ES_{p}(Y)\geq \mathrm{ess}\sup \ES_{p}(X|\W)$.
\end{itemize}
\end{example}
 One can easily check that the functionals $G_{\W}$ of the distortion factor risk measures defined in (i), (ii) and (iv) of Example \ref{example:coherent} satisfy condition A. By Proposition \ref{prop:coherent}, they are all coherent factor risk measures. Moreover, $\mathbb E\left(\ES_{g(\W)}(X|\W)\right)$ is the average value of $\ES$ under different scenarios. In terms of capital requirement, it can be interpreted as: For each scenario $\{\W=\w\}$, the requirement is calculated using $\ES$ at level $g(\w)$, where the level varies for different scenarios.  The overall capital requirement is summarized as an expected value across all scenarios.

 Note that $\ES_{p}(X|\W\in\mathbf B)$ is in fact an extension of the Co$\ES$ and it is not equal to $\mathbb E\left(\ES_{p}(X|\W)\id_{\{\W\in\mathbf B\}}\right)/\mathbb P(\W\in\mathbf B)$.
Note also that for $p,q\in (0,1)$, $\ES_q(\VaR_p(X|\W))$ is a special case of (ii) of Example \ref{exam:distortion}, and  the factor risk measures  $\mathrm{ess}\sup\ES_p(X|\W)$ and $\ES_q(\ES_p(X|\W))$ will be discussed in Section \ref{sec:Coherent} as both of them are coherent factor risk measures.
\section{Quantile factor risk measure}\label{sec:VaR}


Next, we consider the characterization of $\VaR$-type risk measures
in terms of conditional distribution of $X$ on the factors $\W$.  We shall offer expressions with two different parametric systems.
We start with the following notation.

We say $D\subset D^{*}$ is \emph{increasing}
if $1\in D$ and $0\notin D$, and $g\in D$ whenever $g\geq f$ for
some $f\in D$. Let $\mu$ be a probability measure on $(\R^{N},\mathcal{B}(\R^{N}))$. The notation $f\in_{\mu}D$ means there exists
$g\in D$ such that $\mu(\{\w:f(\w)=g(\w)\})=1$. We say a family
of subsets of $D^{*}$, $\{D_{\W}:\W\in\mathcal{Y}\}$, is
law-invariant if  $\W\overset{d}{=}\W'$ implies $D_{\W}=D_{\W'}$ for $\W,\W'\in\mathcal Y$.

Let $S_{\W}$ represent the collection of all $(\A,\BB)$ with $\boldsymbol{\alpha}=(\alpha_{1},\dots,\alpha_{N})\in(0,1)^{N}$
and $\BB=(\beta_{1},\dots,\beta_{N})\in(0,1]^{N}$ such that $\mathbb{P}(\mathrm{VaR}_{\boldsymbol{\alpha}}(\W)\leq\W\leq\mathrm{VaR}_{\BB}(\W))>0$,
where $\mathrm{VaR}_{\boldsymbol{\alpha}}(\W)=(\mathrm{VaR}_{\alpha_{1}}(W_{1}),\dots,\mathrm{VaR}_{\alpha_{N}}(W_{N}))$.
Let $\widehat{D}_{\W}$ represent a set of functions $f:S_{\W}\to[0,1]$
for $\W\in\Y$. We say $\widehat{D}_{\W}$ is increasing
if $0\notin\widehat{D}_{\W}$ and $1\in\widehat{D}_{\W}$, and $g\in\widehat{D}_{\W}$
whenever $g\geq f$ for some $f\in\widehat{D}_{\W}$.  The law-invariance
of $\{\widehat{D}_{\W}:\W\in\mathcal{Y}\}$ is adapted as: For $\W,\W'\in\mathcal Y$, $\W\overset{d}{=}\W'$ implies $\widehat{D}_{\W}=\widehat{D}_{\W'}$.  Let $\mathbb P\circ \W^{-1}(\B)=\mathbb{P}(\W\in \B)$ for $\B\in\mathcal{B}(\R^N)$.
\begin{thm}
\label{thm:MVaR} A mapping $\rho:\mathcal{X}\times\mathcal{Y}\to\mathbb{R}$
satisfies conditions M, OR and LI if and only if one of the following
statements holds:
\begin{enumerate}
\item[(i)] There exists a law-invariant family of increasing sets
$\{D_{\W}:\W\in\mathcal{Y}\}$ such that
\begin{equation}
\rho\left(X,\W\right)=\inf\left\{ x:F_{X|\W=\cdot}(x)\in_{\mathbb{P}\circ\W^{-1}}D_{\W}\right\} .\label{eq:3}
\end{equation}
\item[(ii)] There exists a law-invariant family of increasing sets
$\{\widehat{D}_{\W}:\W\in\mathcal{Y}\}$ such that
\begin{equation}
\rho\left(X,\W\right)=\inf\left\{ x:\left(F_{X|\mathrm{VaR}_{\boldsymbol{\alpha}}(\W)\leq\W\leq\mathrm{VaR}_{\BB}(\W)}(x)\right)_{(\A,\BB)\in S_{\W}}\in\widehat{D}_{\W}\right\} .\label{eq:VaRC}
\end{equation}
\end{enumerate}
\end{thm}
For a family of increasing sets
$\{D_{\W}:\W\in\mathcal{Y}\}$, we denote $\rho$ defined by \eqref{eq:3} by $q_{D_{\W}}(X,\W)$ and we call it \emph{quantile factor risk measure} with level set $D_{\W}$ and factor $\W$. Note that the classical left quantile can be expressed as $\VaR_\alpha(X)=\inf\{x: F_X(x)\in [\alpha,\infty)\},~\alpha\in (0,1]$ and the classical right quantile can be expressed as $\VaR_\alpha^+(X)=\inf\{x: F_X(x)\in (\alpha,\infty)\},~\alpha\in [0,1)$.  Clearly, if $\W$ is a constant, then $q_{D_{\W}}(X,\W)$ boils down to the classical quantiles.  Hence, $q_{D_{\W}}(X,\W)$ can be viewed as an extension of the classical quantiles.

Moreover, $q_{D_{\W}}(X,\W)$ satisfies monotonicity and ordinality, which can be seen from the proof of Theorem \ref{thm:MVaR} in Appendix \ref{Appendix:VaR}.  Similarly for the distortion factor risk measure, the law-invariance of $\{D_{\W}:\W\in\mathcal{Y}\}$ is a sufficient  but not a necessary condition to guarantee the law-invariance of $q_{D_{\W}}(X,\W)$.  A sufficient and necessary condition is the \emph{weak law-invariance} of $\{D_{\W}:\W\in\mathcal{Y}\}$: For $A, A'\in\mathcal F,\W, \W'\in\mathcal Y$ satisfying $(\id_A,\W)\overset{d}{=}(\id_{A'},\W')$, we have either $\mathbb{P}(A|\W=\cdot)\in_{\mathbb P\circ \W^{-1}} D_{\W}\cap D_{\W'}$
or $\mathbb{P}(A|\W=\cdot)\notin_{\mathbb P\circ \W^{-1}} D_{\W}\cup D_{\W'}$.

\begin{prop}\label{VaR:weaklawinvariance}
For a quantile factor risk measure $q_{D_{\W}}(X,\W)$ with a family of increasing sets
$\{D_{\W}:\W\in\mathcal{Y}\}$, it is law-invariant if and only if $\{D_{\W}:\W\in\mathcal{Y}\}$ is weakly law-invariant.
\end{prop}

If $N=1$, then $S_{W}$ in (ii) of Theorem \ref{thm:MVaR} is independent
of $W$: $S_{W}=\{(\alpha,\beta):0<\alpha<\beta\leq1\}$. Hence \eqref{eq:VaRC}
becomes
\[
\rho\left(X,W\right)=\inf\left\{ x:\left(F_{X|\mathrm{VaR}_{\alpha}(W)\leq W\leq\mathrm{VaR}_{\beta}(W)}(x)\right)_{0<\alpha<\beta\leq1}\in\widehat{D}_{W}\right\} .
\]
If we restrict $\W$ as discrete random vectors, (i) of Theorem \ref{thm:MVaR}
admits a simplified expression. Let $\mathcal{Y}_{D}$ be the set of all discrete random vectors $\W$. For $\W\in\Y_{D}$, we denote $S_{\W}^{1}=\{\w=(w_{1},\dots,w_{N}):\mathbb{P}(\W=\w)>0\}.$
 We say $D_{\W}^{1}\subsetneq[0,1]^{S_{\W}^{1}}$ is an \emph{increasing
set} if $\mathbf{1}=(1)_{\w\in S_{\W}^{1}}\in D_{\W}^{1}$, $\mathbf{0}=(0)_{\w\in S_{\W}^{1}}\notin D_{\W}^{1}$
and $\mathbf{y}=(y_{\w})_{\w\in S_{\W}^{1}}\in D_{\W}^{1}$ if $y_{\w}\geq x_{\w},\w\in S_{\W}^{1}$
for some $\mathbf{x}\in D_{\W}^{1}$. The law-invariance of $\{D_{\W}^{1}:\W\in\mathcal{Y}\}$:
For $\W,\W'\in\mathcal{Y}$ satisfying $\W\overset{d}{=}\W'$, we
have $D_{\W}^{1}=D_{\W'}^{1}$. 
\begin{cor}
\label{prop:VaR} A mapping $\rho:\mathcal{X}\times\mathcal{Y}_{D}\to\mathbb{R}$
satisfies conditions M, OR and LI if and only if there exists a law-invariant family
of increasing sets $\{D_{\W}^{1}:\W\in\mathcal{Y}_{D}\}$
such that
\begin{equation}
\rho\left(X,\W\right)=\inf\left\{ x:\left(F_{X|\W=\mathbf{w}}(x)\right)_{\mathbf{w}\in S_{\W}^{1}}\in D_{\W}^{1}\right\} .\label{eq:VaRD}
\end{equation}
\end{cor}


If $\mathcal Y$ is a set of constants, then  Theorem \ref{thm:MVaR} boils down to the result in \cite{C09} and \cite{FLW22}, which offers the characterization of the classical quantiles. Hence, Theorem \ref{thm:MVaR} defines a more general quantiles summarizing the quantiles under different scenarios into a single value.

We next consider some special cases of Theorem \ref{thm:MVaR} and
Corollary \ref{prop:VaR}.
\begin{example}\label{example3}
In some special cases, \eqref{eq:3} in Theorem \ref{thm:MVaR} has
the following simplified representations.
\begin{enumerate}
\item[(i)] For $q\in (0,1)$ and $g\in D^*$ with $\mathbb P(0<g(\W)<1)=1$, let $D_{\W}=\{f\in D^{*}: \mathbb P(f(\W)\geq g(\W))\geq q\}$, then
\begin{align*}q_{D_{\W}}(X,\W)&=\inf\left\{ x: \mathbb P(F_{X|\W}(x)\geq g(\W))\geq q\right\}\\
&=\inf\left\{ x: \mathbb P(\VaR_{g(\W)}(X|\W)\leq x)\geq q\right\}=\VaR_q(\VaR_{g(\W)}(X|\W)).\end{align*}
If $q=1$, then \begin{align*}q_{D_{\W}}(X,\W)=\mathrm{ess}\sup\VaR_{g(\W)}(X|\W).
\end{align*}
\item[(ii)] For $p,q\in (0,1)$, let $D_{\W}=\{f\in D^{*}: \mathbb P(f(\W)\geq p)\geq q\}$, then
\begin{align*}q_{D_{\W}}(X,\W)=\VaR_q(\VaR_p(X|\W)).
\end{align*}
If $q=1$, then \begin{align*}q_{D_{\W}}(X,\W)=\mathrm{ess}\sup\VaR_{p}(X|\W).\end{align*}
\end{enumerate}
\end{example}
For $\VaR_q(\VaR_{g(\W)}(X|\W))$, it can be interpreted in the context of regulatory capital requirement as below.  For each scenario $\{\W=\w\}$, the capital requirement is calculated using $\VaR$ at level $g(\w)$, where the level $g$ varies with respect to different scenarios. The overall capital  requirement is summarized by another $\VaR$ to cover the capital requirement  for $100 q\%$ different scenarios. For $q=1$, $\mathrm{ess}\sup\VaR_{g(\W)}(X|\W)$ guarantees the capital requirement is satisfied for all scenarios.

The following two examples are closely related to $\mathrm{Co}\VaR$.
\begin{example}
For $p_{\A,\BB}\in(0,1),(\A,\BB)\in S_{\W}$, \eqref{eq:VaRC} can
be simplified as follows.
\begin{enumerate}
\item[(i)] If $\widehat{D}_{\W}=\prod_{(\A,\BB)\in S_{\W}}[p_{\A,\BB},1]$,
then
\[
\rho\left(X,\W\right)=\sup_{(\A,\BB)\in S_{\W}}\mathrm{VaR}_{p_{\A,\BB}}(X|\mathrm{VaR}_{\boldsymbol{\alpha}}(\W)\leq\W\leq\mathrm{VaR}_{\BB}(\W));
\]
\item[(ii)] If $(\A_{0},\BB_{0})\in S_{\W}$ and $\widehat{D}_{\W}=(\prod_{(\A,\BB)\in S_{\W}\setminus\{(\A_{0},\BB_{0})\}}[0,1])\times[p_{\A_{0},\BB_{0}},1]$,
then
\[
\rho\left(X,\W\right)=\mathrm{VaR}_{p_{\A_{0},\BB_{0}}}(X|\mathrm{VaR}_{\boldsymbol{\alpha}_{0}}(\W)\leq\W\leq\mathrm{VaR}_{\BB_{0}}(\W));
\]
\item[(iii)] If additionally $\BB_{0}=\mathbf{1}$ for (ii), then
\[
\rho\left(X,\W\right)=\mathrm{VaR}_{p_{\A_{0},\BB_{0}}}(X|\W\geq\mathrm{VaR}_{\boldsymbol{\alpha}_{0}}(\W)).
\]
\end{enumerate}
\end{example}

\begin{example}
For $p_{\w}\in(0,1),\w\in S_{\W}^{1}$, and $\W\in\mathcal{Y}_{D}$,
expression \eqref{eq:VaRD} can be reduced to the following expressions.
\begin{enumerate}
\item[(i)] If $D_{\W}^{1}=\prod_{\w\in S_{\W}^{1}}[p_{\w},1]$, then
\[
\rho\left(X,\W\right)=\sup_{\boldsymbol{\w}\in S_{\W}^{1}}\mathrm{VaR}_{p_{\w}}(X|\W=\w);
\]
\item[(ii)] If $\w_{0}\in S_{\W}^{1}$ and $D_{\W}^{1}=(\prod_{\w\in S_{\W}^{1}\setminus\{\w_{0}\}}[0,1])\times[p_{\w_{0}},1]$,
then
\[
\rho\left(X,\W\right)=\mathrm{VaR}_{p_{\w_{0}}}(X|\W=\w_{0}).
\]
\end{enumerate}
\end{example}


\section{Linear factor risk measures}\label{Sec:linear}

\label{sec:linear} In this section, we consider linear factor risk
measures, which includes MES as a special
case. We say a family of probability measures on $(\R^{N},\mathcal{B}(\R^{N}))$,
$\{Q_{\W}:\W\in\Y\}$, is \emph{law-invariant} if $Q_{\W}=Q_{\W'}$
whenever $\W\overset{d}{=}\W'$ for $\W,\W'\in\mathcal{Y}$; we say $\rho:\X\times \mathcal Y\to\R$ is \emph{additive} (AD) if $\rho(X+Y,\W)=\rho(X,\W)+\rho(Y,\W)$ for all $X,Y\in\X$. 
\begin{thm}
\label{thm:MES} A mapping $\rho:\mathcal{X}\times\mathcal{Y}\to\mathbb{R}$
satisfies conditions M, AD, IC, N and LI if and only if there exists
a law-invariant family of probability measures $\{Q_{\W}:\W\in\Y\}$
on $(\R^{N},\mathcal{B}(\R^{N}))$ such that $Q_{\W}<<\mathbb{P}\circ\W^{-1}$,
and
\begin{equation}
\rho(X,\W)=E^{Q_{\W}}(\mathbb{E}(X|\W=\cdot))=\int_{\R^{N}}\mathbb{E}(X|\W=\w)\d Q_{\W}.\label{eq:MES}
\end{equation}
\end{thm}

Note that if $Q_{\W}=\mathbb{P}\circ\W^{-1}$, then \eqref{eq:MES}
boils down to the law of iterated expectation. Hence Theorem \ref{thm:MES}
can be viewed as an extension of the law of iterated expectation.
Moreover, as $\mathbb{E}(X|\W=\w)$ is the so-called conditional expectation, \eqref{eq:MES}
can be demonstrated as the weighted average of the conditional expectation. It will
be more clear for the discrete random vector $\W$. Let us see some
examples as below.
\begin{enumerate}
\item[(i)] If $\mathbb{P}(\W=\w_{i})>0,~i=1,\dots,n$ and $\sum_{i=1}^{n}\mathbb{P}(\W=\w_{i})=1$,
then \eqref{eq:MES} becomes
\[
\rho(X,\W)=\sum_{i=1}^{n}c_{i}\mathbb{E}(X|\W=\w_{i})
\]
with $c_{i}\geq0$ and $\sum_{i=1}^{n}c_{i}=1$. Note that $\rho(X,\W)$ has the same form as Moody's rating measure for credit rating in \cite{M23}.
\item[(ii)] For $\B\in\mathcal{B}(\R^{N})$ with $\mathbb{P}\circ\W^{-1}(\B)>0$,
let $Q_{\W}(\cdot)=\frac{\mathbb{P}\circ\W^{-1}(\cdot\cap\B)}{\mathbb{P}\circ\W^{-1}(\B)}$.
Then \eqref{eq:MES} is reduced to
\[
\rho(X,\W)=\mathbb{E}(X|\W\in\B),
\]
which extends MES.
\end{enumerate}

\section{Coherent factor risk measure}\label{sec:Coherent}
In this section, we aim to find the expression of coherent factor risk measures. We start with an extension of the Hardy-Littewood inequality, which is
 inspired by Proposition 4.6 in Dela Vega and Elliott (2021) and Lemma 3.3 of de Castro et al.(2024).  This extension is crucial to characterize the coherent factor risk measure and it is also of interest independently. For $X\in\X$ and $\W\in\Y$, let $L(X,\W)=\{Z\in\X: (Z,\W)\overset{d}{=}(X,\W)\}$. In this section, we suppose that for any $\W\in\mathcal Y$, there exists $U\sim U(0,1)$ that is independent of $\W$.
\begin{lem}\label{le11} For $X\in\X$, $\W\in\mathcal Y$ and $Y\in L^1$, we have
\begin{align*}\sup_{Z\in L(X,\W)}\mathbb{E}\left(ZY\right)&=\mathbb{E}\left(\int_{0}^{1}\VaR_{t}(X|\W)\VaR_t\left(Y|\W\right)\d t\right),\\
\inf_{Z\in L(X,\W)}\mathbb{E}\left(ZY\right)&=\mathbb{E}\left(\int_{0}^{1}\VaR_{1-t}(X|\W)\VaR_t\left(Y|\W\right)\d t\right).
\end{align*}
\end{lem}
Let us next introduce some notation and properties.
We denote by $\mu_1(\mathbb P)$ the set of all probability measures on $(\Omega, \mathcal F)$ that are absolutely continuous with respect to $\mathbb P$.
We say a family of subsets of $\mu_1(\mathbb P)$, $\{\mathcal{Q}_{\W}:\W\in\mathcal Y\}$, is \emph{law-invariant} if for $\W,\W'\in \mathcal Y$, $\W\overset{d}{=}\W'$ implies
$\{F_{Z, \W}: Z=\frac{\d Q}{\d \mathbb{P}},~Q\in \mathcal{Q}_{\W}\}=\{F_{Z, \W'}: Z=\frac{\d Q}{\d \mathbb{P}},~Q\in \mathcal{Q}_{\W'}\}$.
For a mapping $\rho:\X\times\mathcal Y\to\R$, we say $\rho$ is \emph{continuous from above} if $\rho(X_n,\W)\downarrow \rho(X,\W)$ whenever $X_n\downarrow X$.

In what follows, we characterize the coherent factor risk measure.
\begin{thm}
\label{prop:distortion-distribution-1} A mapping $\rho:\mathcal{X}\times\mathcal{Y}\to\mathbb{R}$ is coherent, continuous from above and law-invariant if and only if there exists a law-invariant family of sets of probability measures $\{\mathcal{Q}_{\W}:\W\in\mathcal Y\}$ such that
\begin{align}\label{coherent}
\rho\left(X,\W\right)=\sup_{Q\in \mathcal{Q}_{\W}}\mathbb{E}\left(\int_{0}^{1}\VaR_{t}(X|\W)\VaR_t\left(\frac{\d Q}{\d \mathbb P}\Big|\W\right)\d t\right).
\end{align}
\end{thm}

For the case of discrete $\W$ taking only finite values, Theorem \ref{prop:distortion-distribution-1}  boils down to Theorem 3.8 of \cite{WZ21}. This can be seen from the following result.
\begin{cor}\label{cor:coherent} Under the assumption of Theorem \ref{prop:distortion-distribution-1}, if $\mathcal Y$ is a set of discrete random variables, then \eqref{coherent} is reduced to
$$
\rho\left(X,\W\right)=\sup_{Q\in \mathcal{Q}_{\W}}\sum_{\w\in S_{\W}^1}\mathbb P(\W=\w)\int_{0}^{1}\VaR_{t}(X|\W=\w)\VaR_t\left(\frac{\d Q}{\d \mathbb P}\Big|\W=\w\right)\d t.
$$
\end{cor}
Alternatively, we can construct coherent factor risk measures as follows.
Using  Proposition 11.9 of \cite{FS16}, we immediately arrive at the following result.
\begin{prop} Let $\varrho:\X\to\R$ be a law-invariant coherent risk measure. Then the mapping   $(X,\W)\mapsto \varrho(\ES_p(X|\W))$ with $(X,\W)\in \mathcal{X}\times\mathcal{Y}$ is a law-invariant coherent factor risk measure.
\end{prop}
\begin{example} For $p,q\in (0,1)$,  $\mathrm{ess}\sup\ES_p(X|\W)$ and $\ES_q(\ES_p(X|\W))$  are two families of law-invariant coherent factor risk measures.
\end{example}

\section{Risk management applications}\label{sec:risksharing}

In this section, we consider the application of our main results to risk sharing problem and the evaluation of the risk with factors.
\subsection{Risk sharing}
In this subsection, we consider the risk sharing problem among multiple agents with the preference represented by the distortion factor risk measures. Risk sharing problem with risk measures has been widely studied in the literature. We refer to \cite{BE05} and \cite{FS08} for risk sharing with convex risk measures,  and \cite{ELW18} and \cite{LMWW22} for risk sharing problem with quantile-based risk measures. Our interest  here is  the comonotonic risk sharing problem, which is widely applied in optimal insurance and reinsurance contract design; see e.g., \cite{A63}, \cite{ABT17} and \cite{CC20} and the references therein.

For $X\in\X$, we denote all the comonotonic allocations by $$\mathbb A_n^+(X)=\{(X_1,\dots, X_n): X_i~ \text{and}~X~ \text{are comonotonic},~i=1,\dots, n, ~\text{and}~\sum_{i=1}^{n}X_i=X\}.$$
 The risk sharing problem is defined as
$$\boxplus_{i=1}^n\rho_i(X,\W_i)=\inf\left\{\sum_{i=1}^{n}\rho_i(X_i,\W_i): (X_1,\dots, X_n)\in \mathbb A_n^+(X)\right\}.$$
We say an allocation $(X_1,\dots, X_n)\in \mathbb A_n^+(X)$ is an \emph{optimal allocation} if $\sum_{i=1}^{n}\rho_i(X_i,\W_i)=\boxplus_{i=1}^n\rho_i(X,\W_i)$.
\begin{prop}\label{prop:risksharing} For continuous from below distortion functionals $G_{\W_i},~i=1,\dots,n$, we have
$$\boxplus_{i=1}^n\rho_{G_{\W_i}}(X,\W_i)=\int_{0}^{\infty} G_{\W_1,\dots,\W_n,X}(x)\d x+\int_{-\infty}^{0}\left(G_{\W_1,\dots,\W_n,X}(x)-1\right)\d x,$$
where $G_{\W_1,\dots,\W_n,X}(x)=\min_{i=1,\dots,n}G_{\W_i}(1-F_{X|\W_i=\cdot}(x))$. Moreover,
 the optimal allocations are $X_i=h_i(X),~i=1,\dots,n$ with $h_i(x)=\int_{0}^{x} r_i(t)\d t$, where $r_i$ are non-negative measurable functions satisfying
 $r_i(x)=0$ if $G_{\W_i}(1-F_{X|\W_i=\cdot}(x))>G_{\W_1,\dots,\W_n,X}(x)$, and $\sum_{i=1}^{n} r_i(x)=1,~x\in\R$.
\end{prop}

Using Proposition \ref{prop:risksharing} and the expressions in Examples \ref{exam:distortion}-\ref{example:coherent}, we immediately arrive at the following results.
\begin{cor}
\begin{enumerate}
\item[(i)] For left-continuous $\Lambda_i\in \mathfrak{D}$ and $p_i\in (0,1)$,
let $\rho_i(X,\W_i)=\varrho_{\Lambda_i}(\VaR_{p_i}(X|\W_i))$. Then we have
\begin{align*}\boxplus_{i=1}^n\rho_i(X,\W_i)&=\int_{0}^{\infty} \min_{i=1,\dots, n}\Lambda_i(\mathbb P(F_{X|\W_i}(x)<p_i))\d x\\
&+\int_{-\infty}^{0}\left(\min_{i=1,\dots, n}\Lambda_i(\mathbb P(F_{X|\W_i}(x)<p_i))-1\right)\d x;
\end{align*}
\item[(ii)] For $p_i\in (0,1)$, let $\rho_i(X,\W_i)=\mathbb E\left(\ES_{p_i}(X|\W_i)\right)$. Then we have
\begin{align*}\boxplus_{i=1}^n\rho_i(X,\W_i)&=\int_{0}^{\infty} \min_{i=1,\dots, n}\mathbb E\left(\frac{(1-F_{X|\W_i}(x))\wedge (1-p_i)}{1-p_i}\right)\d x\\
&+\int_{-\infty}^{0}\left(\min_{i=1,\dots, n}\mathbb E\left(\frac{(1-F_{X|\W_i}(x))\wedge (1-p_i)}{1-p_i}\right)-1\right)\d x;
\end{align*}
\item[(iii)] For $p_i, q_i\in (0,1)$,
let $\rho_i(X,\W_i)=\VaR_{q_i}(\VaR_{p_i}(X|\W_i))$. Then we have
\begin{align*}\boxplus_{i=1}^n\rho_i(X,\W_i)&=\int_{0}^{\infty} \min_{i=1,\dots, n}\id_{\{\mathbb P(F_{X|\W_i}(x)<p_i)>1-q_i\}}\d x\\
&+\int_{-\infty}^{0}\left(\min_{i=1,\dots, n}\id_{\{\mathbb P(F_{X|\W_i}(x)<p_i)>1-q_i\}}-1\right)\d x\\
&=\min_{i=1,\dots, n}\VaR_{q_i}(\VaR_{p_i}(X|\W_i)).
\end{align*}
\end{enumerate}
\end{cor}

\subsection{Evaluation of risk using $\VaR_q(\VaR_p(X|\W))$}
In this subsection, we evaluate the risk using a factor risk measure
$\mathrm{VaR}_{q}\left(\mathrm{VaR}_{p}\left(X\vert\W\right)\right)
$ and then compare it with $\VaR_p(X)$. A standard approach in economics is that we regress $X$, over $\W$
and use the model. Hence, let us consider the following model:
\[
X=\beta_{0}+\boldsymbol{\beta}\W+\sigma\epsilon,
\]
where $\epsilon$ is an independent idyosyncratic risk with standard normal
distribution, $\sigma>0$ and $\boldsymbol{\beta}=(\beta_1,\dots, \beta_N)$.  By implementing this into our risk measure, we have
\begin{align*}
\rho\left(X,\W\right) & =\beta_{0}+\mathrm{VaR}_{q}\left(\mathrm{VaR}_{p}\left(\boldsymbol{\beta}\W+\sigma\epsilon\vert\W\right)\right)\\
 & =\beta_{0}+\mathrm{VaR}_{q}\left(\boldsymbol{\beta}\W\right)+\sigma\mathrm{VaR}_{p}\left(\epsilon\right)=\beta_{0}+\mathrm{VaR}_{q}\left(\boldsymbol{\beta}\W\right)+\sigma N^{-1}\left(p\right),
\end{align*}
where $N$ is the CDF of the standard normal random variable.
For non-factor risk measure, we consider
\[
\rho\left(X\right)=\mathrm{VaR}_{p}\left(\beta_{0}+\boldsymbol{\beta}\W+\sigma\epsilon\right).
\]
For each scenario $\{\W=\boldsymbol{w}\}$,
$p$ represents the confidence level of the capital requirement and
$q$ represents the overall emphasis we want to put on the systematic
risk.
From the capital requirement point of view $p=0.95$ or $0.975$,
are standard choices, however, for $q$ we can have a wide range from
$q=0.5$ to $q=0.99$. Let us check the percentage change of $\VaR_q(\VaR_p(X|\W))$ compared to $\VaR_p(X)$, i.e.,
\[
\text{Diff}=\rho\left(X,\W\right)/\rho\left(X\right)-1
=\frac{\VaR_p(\sigma\epsilon)-\mathrm{VaR}_{p}\left(\boldsymbol{\beta}\W-
\mathrm{VaR}_{q}\left(\boldsymbol{\beta}\W\right)+\sigma\epsilon\right)}
{\mathrm{VaR}_{p}\left(\beta_{0}+\boldsymbol{\beta}\W+\sigma\epsilon\right)}.
\]
 Note that the numerator can be interpreted as the risk contribution of the $q$-centralized factors i.e.,
$\boldsymbol{\beta}\W-\mathrm{VaR}_{q}\left(\boldsymbol{\beta}\W\right)$.

\subsubsection{Data}

Here we describe the data set we use in this chapter. The data are
on a monthly basis: a set of 642 months, from February 1952 to August
2012.
We use almost the same data as in \cite{FRW03}.
The data consists of seven different securities, which work as the factors in our risk evaluation:
\begin{enumerate}
\item Risk Free, three-month T-bill rate (as a proxy), denoted by\textit{
(RF)};
\item Market , Market Risk minus Risk Free, denoted by \textit{(RM-RF)};
\item Size, Small Minus Big, denoted by \textit{(SMB)};
\item Book-to-market value, High Minus Low, denoted by \textit{(HML)};
\item Momentum, Up Minus Down, denoted by \textit{(UMD)};
\item Term factor, the difference between a long-term government bond return
and the three-month T-bill rate, denoted by \textit{(TERM)};
\item Default factor, the difference between the return on a portfolio of
long-term corporate bonds and a long-term government bond return,
denoted by \textit{(DEF)}.
\end{enumerate}
Items 1,2,3,4,5 are taken from the Fama and French Library. Items
2,3,4 are the usual three Fama and French factors in \cite{FF92}
and item 5 is used in \cite{C97}. The first five factors
explain the premiums on stocks. Factors 6 and 7 are known as bond-market
factors. The TERM factor is the difference between long-term government
bond, provided by BGFRS\footnote{Board of Governors of the Federal Reserve System},
and the short-term government or Treasury Bonds (T-Bill) which is
the same as risk free. As for DEF, we took the Moody's Aaa rated corporation
bonds, provided by BGFRS.


In  FRED\footnote{Federal Reserve Economic Data}, it has a set of six macro-economic risk variables. We consider the following two risk variables as the risk that will be evaluated later:
\begin{enumerate}
\item Real Interest rate, the monthly return on a three-month T-bill, denoted
by \textit{(RI)};
\item Dividend yield, the monthly dividend yield on the S\&P 500, denoted
by \textit{(DIV)}.
\end{enumerate}
\subsubsection{Numerical results}

\begin{figure}
\subfloat[\small The heatmap for T-bill]{\begin{centering}
\includegraphics[scale=0.5]{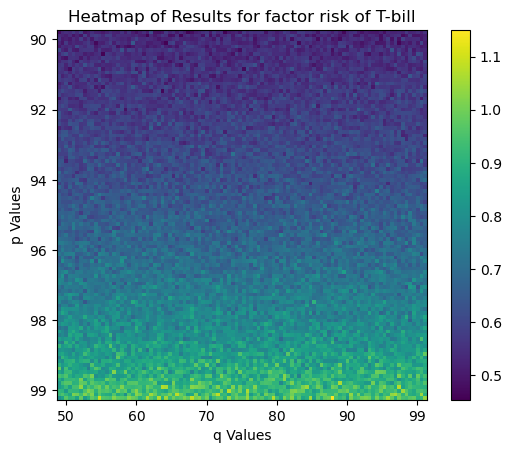}
\par\end{centering}
}\subfloat[\small The heatmap for Dividend]{\begin{centering}
\includegraphics[scale=0.5]{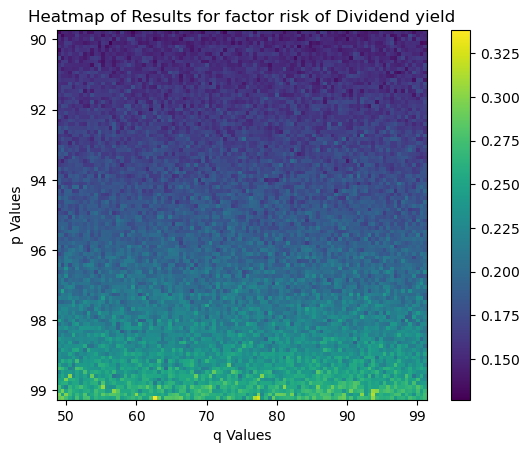}
\par\end{centering}
}
\caption{The heatmaps for $\VaR_q(\VaR_p(X|\W))$}\label{figure:factor value}
\end{figure}

\begin{figure}
\subfloat[\small The heatmap of Diff(T-bill)]{\begin{centering}
\includegraphics[scale=0.5]{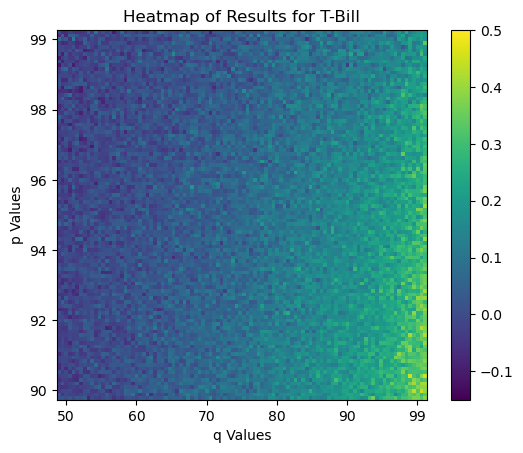}
\par\end{centering}
}\subfloat[\small The heatmap of Diff(Dividend)]{\begin{centering}
\includegraphics[scale=0.5]{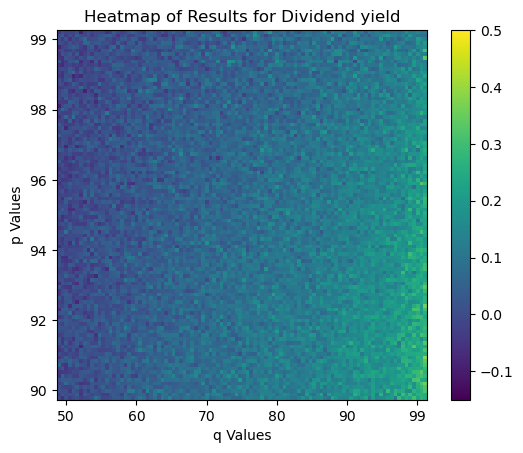}
\par\end{centering}
}
\caption{The heatmaps for Diff}\label{figure:Diff}
\end{figure}

In the following, we set the range of $p$ as $[0.95,0.99]$ and the range of $q$ as $[0.5,0.99]$ and consider economic risk. We present the heatmaps  of the value of $\VaR_q(\VaR_p(X|\W))$ in Figure \ref{figure:factor value} and the heatmaps of Diff to show the
 difference between $\VaR_q(\VaR_p(X|\W))$ and $\VaR_p(X)$ in percentage in Figure \ref{figure:Diff}. The regression model is obtained in Table \ref{table}.

\begin{table}[H]
\centering \caption{Standard Regression Results of T- bill and Dividend yield}\label{table}

\centering{}%
\begin{tabular}{llcccccc}
\toprule
 &  & \textbf{coef} & \textbf{std err} & \textbf{t} & \textbf{P$>|$t$|$} & \textbf{{[}0.025} & \textbf{0.975{]}}\tabularnewline
\midrule
\multirow{8}{*}{T-bill} & \textbf{const} & 0.1413 & 0.061 & 2.322 & 0.021 & 0.022 & 0.261\tabularnewline
 & \textbf{RF} & -0.2686 & 0.124 & -2.161 & 0.031 & -0.513 & -0.025\tabularnewline
 & \textbf{RM-RF} & -0.0155 & 0.004 & -4.021 & 0.000 & -0.023 & -0.008\tabularnewline
 & \textbf{SMB} & -0.0051 & 0.006 & -0.916 & 0.360 & -0.016 & 0.006\tabularnewline
 & \textbf{HML} & -0.0207 & 0.006 & -3.433 & 0.001 & -0.033 & -0.009\tabularnewline
 & \textbf{UMD} & -2.031e-05 & 0.004 & -0.005 & 0.996 & -0.008 & 0.008\tabularnewline
 & \textbf{TERM} & 0.0303 & 0.013 & 2.265 & 0.024 & 0.004 & 0.057\tabularnewline
 & \textbf{DEF} & 0.1532 & 0.033 & 4.694 & 0.000 & 0.089 & 0.217\tabularnewline
\midrule
\multirow{8}{*}{Dividend yield} & \textbf{const} & 0.0325 & 0.017 & 1.891 & 0.059 & -0.001 & 0.066\tabularnewline
 & \textbf{RF} & -0.1273 & 0.035 & -3.631 & 0.000 & -0.196 & -0.058\tabularnewline
 & \textbf{RM-RF} & 0.0022 & 0.001 & 2.060 & 0.040 & 0.000 & 0.004\tabularnewline
 & \textbf{SMB} & -0.0002 & 0.002 & -0.124 & 0.901 & -0.003 & 0.003\tabularnewline
 & \textbf{HML} & -0.0005 & 0.002 & -0.299 & 0.765 & -0.004 & 0.003\tabularnewline
 & \textbf{UMD} & 0.0025 & 0.001 & 2.293 & 0.022 & 0.000 & 0.005\tabularnewline
 & \textbf{TERM} & 0.0119 & 0.004 & 3.159 & 0.002 & 0.005 & 0.019\tabularnewline
 & \textbf{DEF} & 0.0369 & 0.009 & 4.005 & 0.000 & 0.019 & 0.055\tabularnewline
\end{tabular}
\end{table}
In Figure \ref{figure:factor value}, we present the values of $\VaR_q(\VaR_p(X|\W))$ of T-Bill and Dividend for different values of $p$ and $q$.
In Figure \ref{figure:Diff}, we present the Diff(T-Bill) and Diff(Dividend) for different values of $p$ and $q$.

As it is clear that for larger values of $q$, Diff(T-Bill) and Diff(Dividend) are
more positive, which indicates for larger $q$ the value of the factor
risk measure is larger. For smaller $q$, Diff(T-Bill) and Diff(Dividend) become
more negative, which indicates the value of the factor risk measure is smaller than $\VaR_p(X)$ for smaller $q$. We can expect that there is a $q_0\in (0,1)$  such that $\VaR_{q_0}(\VaR_p(X|\W))=\VaR_p(X)$. This means that $\VaR_p(X)$ can satisfy the capital requirement for $100q_0\%$ of different scenarios. We can also see that $\VaR_{q}(\VaR_p(X|\W))$ is more flexible than $\VaR_p(X)$ as it has one additional parameter to adjust the values.

While the values of Diff for T-bill ranges in $[-0.13,0.5]$, Diff
for Dividend yield ranges in $[-0.1,0.38]$. This shows that the contribution
of the factors to the risk causes larger percentage change for T-bill
compared with Dividend yield.

\section{Concluding remarks}

In this paper we motivated and introduced factor risk measures, to
evaluate the risk with regards to a factor vector. The main contribution of this paper is
 to characterize the distortion, quantile, linear and coherent factor risk measures by deriving the explicit expressions for the factor risk measures satisfying some desirable properties.  We have introduced many concrete examples of factor risk measures such as $\varrho_{\Lambda}(\VaR_p(X|\W))$, $\mathbb E(\VaR_{p}(X|\W))$, $\VaR_q(\VaR_p(X|\W))$, $\mathrm{ess}\sup\VaR_p(X|\W)$, $\mathbb E\left(\ES_{p}(X|\W)\right)$, $\mathrm{ess}\sup\ES_p(X|\W)$ and $\ES_q(\ES_p(X|\W))$. Those factor risk measures have potential to be applied in quantitative risk management and other fields, and need further investigation.  We have shown how distortion factor risk measures
can be naturally applied in risk-sharing and how quantile factor risk measures are applied in risk evaluation.


\vspace{0.2cm}

\section{Appendix}

\label{sec:Appendix} In this section, all the proofs of the results
in Sections \ref{sec:preliminary}-\ref{sec:Coherent} will be displayed.
\subsection{Proof of results in Section \ref{sec:preliminary}}
{\bf Proof of Proposition \ref{law-invariance}}.
For $X,Y\in \X$, clearly, $(X,\W)\overset{d}{=}(Y,\W)$ implies $X\overset{d}{=}Y$ under $Q$ for all $Q\in\mathcal Q_{\W}$. Conversely, $X\overset{d}{=}Y$ under $Q$ for all $Q\in\mathcal Q_{\W}$ implies that for $A\in\mathcal B(\R)$, $$\mathbb P\left(\{X\in A\}\cap \{\W\in \prod_{i=1}^N (a_i,b_i)\}\right)=\mathbb P\left(\{Y\in A\}\cap \{\W\in \prod_{i=1}^N (a_i,b_i)\}\right)$$
for all $a_i,b_i\in\mathbb Q$. Note that $\left\{\prod_{i=1}^N (a_i,b_i): a_i,b_i\in\mathbb Q\right\}$ is a $\pi$-system and it can generate $\mathcal B(\R^N)$. Hence, applying Dynkin's $\pi$-$\lambda$ theorem (e.g., Theorem 3.2 of \cite{B95}), we have $\mathbb P\left(\{X\in A\}\cap \{\W\in \B\right)=\mathbb P\left(\{Y\in A\}\cap \{\W\in \B\}\right)$ for all $\B\in \mathcal B(\R^N)$. This implies $(X,\W)\overset{d}{=}(Y,\W)$.  \qed

\subsection{Proof of results in Section \ref{sec:distortion}}\label{Appendix:dis}
Let us first show a new version of the characterization of Choquet integral
in \cite{Sch86}, which plays an important role in the
proofs of Theorems \ref{thm:choquet}-
\ref{thm:MVaR} and Proposition \ref{VaR:weaklawinvariance} later. We say $c:\mathcal{F}\to\mathbb{R}$
is a capacity if $c$ is monotone: $c(A)\leq c(B)$ for $A\subset B$,
and $0=c(\emptyset)\leq c(\Omega)<\infty$.  We say $\rho:\X\times \mathcal Y\to\R$ is \emph{comonotonic monotone (CM)} if for any two comonotonic random
variables $X,Y\in\X$ satisfying $X\le Y$, and any factor $\W\in\Y$, we have $\rho\left(X,\W\right)\le\rho\left(Y,\W\right)$.
\begin{prop}
\label{le1} For a mapping $\rho:\mathcal{X}\times\mathcal{Y}\to\mathbb{R}$,
it satisfies CM and CA if and only if $\rho$ admits the following
expression:
\begin{equation}
\rho(X,\W)=\int_{0}^{\infty}c^{\W}(X>x)\d x+\int_{-\infty}^{0}(c^{\W}(X>x)-c^{\W}(\Omega))\d x,\label{choquet}
\end{equation}
where $c^{\W}:\mathcal{F}\to\mathbb{R}$ is a capacity. Moreover,
$\rho$ additionally satisfies LI if and only if $c^{\W}(A)=c^{\W'}(A')$
whenever $(\mathbb{I}_{A},\W)\overset{d}{=}(\mathbb{I}_{A'},\W')$.
As a by-product, we have that CM and CA is equivalent to M and CA.
\end{prop}

\textbf{Proof of Proposition \ref{le1}}. Note that the "if" part
is obvious. We next fix a $\W\in\mathcal{Y}$ to show the "only if"
part. By CA, we have $\rho(\frac{m}{n}X,\W)=\frac{m}{n}\rho(X,\W)$
for $X\in\mathcal{X}$ and $n,m\in\mathbb{N}$. For $r>0$, there
exist two sequence $\frac{m_{k}}{n_{k}}\uparrow r$ and $\frac{m_{k}'}{n_{k}'}\downarrow r$
as $k\to\infty$. For $X\geq0$, by CM, we have
\[
\frac{m_{k}}{n_{k}}\rho(X,\W)=\rho\left(\frac{m_{k}}{n_{k}}X,\W\right)\leq\rho(rX,\W)\leq\rho\left(\frac{m_{k}'}{n_{k}'}X,\W\right)=\frac{m_{k}'}{n_{k}'}\rho(X,\W).
\]
Letting $k\to\infty$, we obtain $\rho(rX,\W)=r\rho(X,\W)$. One can
similarly show that $\rho(rX,\W)=r\rho(X,\W)$ for $r<0$ and for $X\leq0$
and $r\in\mathbb{R}$. Note that $X1_{\{X>0\}}$ and $X1_{\{X\leq0\}}$
are comonotonic. Hence for $X\in\mathcal{X}$,
\begin{align*}
\rho(rX,\W) & =\rho(rX1_{\{X>0\}},\W)+\rho(rX1_{\{X\leq0\}},\W)\\
 & =r(\rho(X1_{\{X>0\}},\W)+\rho(X1_{\{X\leq0\}},\W))=r\rho(X,\W).
\end{align*}
By Proposition 1 of Schmeidler (1986), we have for $X$ taking only
finite number of values
\[
\rho(X,W)=\int_{0}^{\infty}c^{\W}(X>x)\d x+\int_{-\infty}^{0}(c^{\W}(X>x)-c^{\W}(\Omega))\d x,
\]
where $c^{\W}(A)=\rho(\mathbb{I}_{A},\W)$. Note that $A\subset B$
implies that $\mathbb{I}_{A}\leq\mathbb{I}_{B}$, and $\mathbb{I}_{A}$
and $\mathbb{I}_{B}$ are comonotonic. Hence $c^{\W}(A)=\rho(\mathbb{I}_{A},\W)\leq\rho(\mathbb{I}_{B},\W)=c^{\W}(B)$.
Moreover, $c^{\W}(\emptyset)=\rho(0,\W)=0$. We next show that the
above expression holds for all $X\in\mathcal{X}$. Let $X_{n}=\frac{\lfloor nX\rfloor}{n}$,
where $\lfloor\cdot\rfloor$ represents the floor function. Note that
$X_{n}$ and $X$ are comonotonic and $X_{n}\leq X\leq X_{n}+\frac{1}{n}$.
Hence $\rho(X_{n},\W)\leq\rho(X,\W)\leq\rho(X_{n},\W)+\frac{\rho(1,\W)}{n}$.
This implies
\begin{align*}
\rho(X,\W) & =\lim_{n\to\infty}\rho(X_{n},\W)\\
 & =\lim_{n\to\infty}\int_{0}^{\infty}c^{\W}(X_{n}>x)\d x+\lim_{n\to\infty}\int_{-\infty}^{0}(c^{\W}(X_{n}>x)-c^{\W}(\Omega))\d x\\
 & =\int_{0}^{\infty}c^{\W}(X>x)\d x+\int_{-\infty}^{0}(c^{\W}(X>x)-c^{\W}(\Omega))\d x.
\end{align*}
We establish the first claim. For the second claim, note that LI of
$\rho$ and $c^{\W}(A)=\rho(\mathbb{I}_{A},\W)$ imply $c^{\W}(A)=c^{\W'}(A')$
whenever $(\mathbb{I}_{A},\W)\overset{d}{=}(\mathbb{I}_{A'},\W')$.
Now we show the converse conclusion. For $(X,\W)\overset{d}{=}(X',\W')$,
let $X_{n}=\frac{\lfloor nX\rfloor}{n}$ and $X_{n}'=\frac{\lfloor nX'\rfloor}{n}$.
Then $X_{n}\leq X\leq X_{n}+\frac{1}{n}$ and $X_{n}'\leq X'\leq X_{n}'+\frac{1}{n}$.
We can write $X_{n}=\sum_{i=0}^{m_{n}}x_{i}\mathbb{I}_{A_{i}}$ and
$X_{n}'=\sum_{i=0}^{m_{n}}x_{i}\mathbb{I}_{A_{i}'}$, where $A_{0}=A_{0}'=\Omega$,
$A_{i}\supsetneq A_{i+1}$ and $A_{i}'\supsetneq A_{i+1}'$, $x_{0}\in\mathbb{R}$
and $x_{i}>0,~i=1,\dots,n$. Without loss of generality, we suppose
$x_{0}\geq0$. Applying \eqref{choquet} and noting that $(\mathbb{I}_{A_{i}},\W)\overset{d}{=}(\mathbb{I}_{A_{i}'},\W')$,
we have
\begin{align*}
\rho(X_{n},\W)=\sum_{i=0}^{m_{n}}x_{i}c^{\W}(A_{i})=\sum_{i=0}^{m_{n}}x_{i}c^{\W'}(A_{i}')=\rho(X_{n}',\W').
\end{align*}
Moreover, it follows from \eqref{choquet} that $\rho(X_{n},\W)\leq\rho(X,\W)\leq\rho(X_{n},\W)+\frac{\rho(1,\W)}{n}$
and $\rho(X_{n}',\W')\leq\rho(X',\W')\leq\rho(X_{n}',\W')+\frac{\rho(1,\W')}{n}$.
Hence, we obtain $\rho(X,\W)=\rho(X',\W')$ by letting $n\to\infty$.
We establish the second claim. \qed \vspace{0.2cm}

\vspace{0.2cm}

\textbf{Proof of Theorem \ref{thm:choquet}}. We first show the "if"
part. If $X\leq Y$, then for any $x\in\R$, we have $F_{X|\W=\cdot}(x)\geq_{\mathbb{P}\circ\W^{-1}}F_{Y|\W=\cdot}(x)$.
It follows from the monotonicity of $G_{\W}$ that
\[
G_{\W}\left(1-F_{X|\W=\cdot}(x)\right)\leq G_{\W}\left(1-F_{Y|\W=\cdot}(x)\right)
\]
for all $x\in\R$, leading to $\rho(X,\W)\leq\rho(Y,\W)$. Hence, M of
$\rho$ is verified. For two comonotonic random variables $X$ and
$Y$, there exist increasing and Lipschitz continuous functions $\phi_{1}$
and $\phi_{2}$ satisfying $\phi_{1}(x)+\phi_{2}(x)=x,~x\in\R$ such
that $X=\phi_{1}(X+Y)$ and $Y=\phi_{2}(X+Y)$. It follows that
\begin{align*}
\rho(X,\W) & =\int_{0}^{\infty}G_{\W}\left(1-F_{\phi_{1}(X+Y)|\W=\cdot}(x)\right)\d x+\int_{-\infty}^{0}\left(G_{\W}\left(1-F_{\phi_{1}(X+Y)|\W=\cdot}(x)\right)-1\right)\d x\\
 & =\int_{0}^{\infty}G_{\W}\left(1-F_{X+Y|\W=\cdot}(\phi_{1}^{-1}(x))\right)\d x+\int_{-\infty}^{0}\left(G_{\W}\left(1-F_{X+Y|\W=\cdot}(\phi_{1}^{-1}(x))\right)-1\right)\d x\\
 & =\int_{0}^{\infty}G_{\W}\left(1-F_{X+Y|\W=\cdot}(x)\right)\d\phi_{1}(x)+\int_{-\infty}^{0}\left(G_{\W}\left(1-F_{X+Y|\W=\cdot}(x)\right)-1\right)\d\phi_{1}(x),
\end{align*}
where $\phi_{1}^{-1}(x)=\inf\{y:\phi_{1}(y)>x\}$ with the convention
that $\inf\emptyset=\infty$. Analogously, we have
\begin{align*}
\rho(Y,\W)=\int_{0}^{\infty}G_{\W}\left(1-F_{X+Y|\W=\cdot}(x)\right)\d\phi_{2}(x)+\int_{-\infty}^{0}\left(G_{\W}\left(1-F_{X+Y|\W=\cdot}(x)\right)-1\right)\d\phi_{2}(x).
\end{align*}
Hence $\rho(X,\W)+\rho(Y,\W)=\rho(X+Y,\W)$, meaning that $\rho$
satisfies CA. Direct computation shows
\[
F_{1|\W=\cdot}(x)=_{\mathbb{P}\circ W^{-1}}\left\{ \begin{array}{cc}
0, & x<1\\
1, & x\geq1
\end{array}\right.,
\]
which implies
\[
\rho(X,\W)=\int_{0}^{1}G_{\W}\left(1\right)\d x+\int_{1}^{\infty}G_{\W}(0)\d x+\int_{-\infty}^{0}\left(G_{\W}\left(1\right)-1\right)\d x=1.
\]
Hence N of $\rho$ is satisfied. Finally, we show LI of $\rho$. For
$X,X'\in\X$ and $\W,\W'\in\mathcal{Y}$ satisfying $(X,\W)\overset{d}{=}(X',\W')$,
we have for any $x\in\R$, $(\id_{\{X\leq x\}},\W)\overset{d}{=}(\id_{\{X'\leq x\}},\W')$.
It follows from the law-invariance of $\{G_{\W}:\W\in\mathcal{Y}\}$
that $G_{\W}(1-\mathbb{P}(X\leq x|\W=\cdot))=G_{\W'}(1-\mathbb{P}(X'\leq x|\W'=\cdot))$.
Hence by \eqref{eq:quantiles}, we have $\rho(X,\W)=\rho(X',\W')$.
The LI of $\rho$ is verified.

We next show the "only if" part. In light of Proposition \ref{le1},
M, CA, N and LI imply that $\rho$ has the following representation:
\begin{equation}
\rho(X,\W)=\int_{0}^{\infty}c^{\W}(X>x)\d x+\int_{-\infty}^{0}(c^{\W}(X>x)-1)\d x,\label{choquet11}
\end{equation}
where $c^{\W}$ is a capacity with $c^{\W}(\Omega)=1$ satisfying
$c^{\W}(A)=c^{\W'}(A')$ whenever $(\mathbb{I}_{A},\W)\overset{d}{=}(\mathbb{I}_{A'},\W')$.
For $f\in D^{*}$, define $G_{\W}(f)=\sup\{c^{\W}(A):A\in\mathcal{F}~\text{such that}~f\geq_{\mathbb{P}\circ\W^{-1}}\mathbb{P}(A|\W=\cdot)\}$.
Direct computation gives $G_{\W}(0)=\sup\{c^{\W}(A):\mathbb{P}(A)=0\}=0$
and $G_{\W}(1)=\sup\{c^{\W}(A):A\in\mathcal{F}\}=1$. Moreover, it
follows from the definition that $f\geq_{\mathbb{P}\circ\W^{-1}}g$
implies $G_{\W}(f)\geq G_{\W}(g)$. Hence $G_{\W}$ is monotone over
$D^{*}$.

Next, we show that $G_{\W}(\mathbb{P}(A|\W=\cdot))=c^{\W}(A)$ for
all $A\in\mathcal{F}$. By the definition, we have $G_{\W}(\mathbb{P}(A|\W=\cdot))\geq c^{\W}(A)$.
For any $B\in\mathcal{F}$ satisfying $\mathbb{P}(A|\W=\cdot)\geq_{\mathbb{P}\circ\W^{-1}}\mathbb{P}(B|\W=\cdot)$,
we have $\mathbb{P}(A|\W)\geq\mathbb{P}(B|\W)$ a.s.. Note that there
exist $\W'$ and $U$ such that $\W'\overset{d}{=}\W$ and $U\sim U[0,1]$
is independent of $\W'$. Then $(F_{\mathbb{I}_{A}|\W'}^{-1}(U),\W')\overset{d}{=}(\mathbb{I}_{A},\W)$
and $(F_{\mathbb{I}_{B}|\W'}^{-1}(U),\W')\overset{d}{=}(\mathbb{I}_{B},\W)$,
where $F_{X|\boldsymbol{w}}^{-1}$ is the quantile function of $F_{X|\W=\boldsymbol{w}}$.
Note that $F_{\mathbb{I}_{A}|\W'}^{-1}(U)\geq F_{\mathbb{I}_{B}|\W'}^{-1}(U)$
a.s., and both of them are indicator functions. We denote $A'=\{F_{\mathbb{I}_{A}|\W'}^{-1}(U)=1\}$
and $B'=\{F_{\mathbb{I}_{B}|\W'}^{-1}(U)=1\}$. Then we have $B'\subset A'$
a.s. and $F_{\mathbb{I}_{A}|\W'}^{-1}(U)=\mathbb{I}_{A'},F_{\mathbb{I}_{B}|\W'}^{-1}(U)=\mathbb{I}_{B'}$
a.s.. By LI, we have $c^{\W}(A)=\rho(\mathbb{I}_{A},\W)=\rho(\mathbb{I}_{A'},\W')$
and $c^{\W}(B)=\rho(\mathbb{I}_{B},\W)=\rho(\mathbb{I}_{B'},\W')$.
By M, we have $c^{\W}(A)\geq c^{\W}(B)$, which implies $G_{\W}(\mathbb{P}(A|\W=\cdot))\leq c^{\W}(A)$.
Hence $G_{\W}(\mathbb{P}(A|\W=\cdot))=c^{\W}(A)$ for all $A\in\mathcal{F}$. Replacing $c^{\W}(X>x)$ by $G_{\W}(\mathbb{P}(X>x|\W=\cdot))$ in \eqref{choquet11}, we obtain \eqref{eq:quantiles}. Note that there exist $\W'\in\mathcal Y$ and $U\sim U(0,1)$ such that $\W'\overset{d}{=}\W$ and $U$ is independent of $\W'$. Hence, for any $X\in\X$, there exists $X'\in \X$ such that $(X,\W)\overset{d}{=}(X',\W')$.
By LI of $\rho$, we have \begin{align*}\rho(X,\W)&=\rho(X',\W')\\
&=\int_{0}^{\infty}G_{\W'}\left(1-F_{X'|\W'=\cdot}(x)\right)\d x+\int_{-\infty}^{0}\left(G_{\W'}\left(1-F_{X'|\W'=\cdot}(x)\right)-1\right)\d x\\
&=\int_{0}^{\infty}G_{\W'}\left(1-F_{X|\W=\cdot}(x)\right)\d x+\int_{-\infty}^{0}\left(G_{\W'}\left(1-F_{X|\W=\cdot}(x)\right)-1\right)\d x.
\end{align*}
This implies that $G_{\W}$ can be chosen as $G_{\W}=G_{\W'}$. Hence, we can choose a law-invariant family of $\{G_{\W}:\W\in\Y\}$ such that \eqref{eq:quantiles} holds.
 \qed

\vspace{0.2cm}
{\bf Proof of Proposition \ref{weakly law-invariant}}. We first show the "if" part. For $X, X'\in\X$ and $\W,\W'\in\mathcal Y$ satisfying  $(X,\W)\overset{d}{=}(X',\W')$, we have for any $x\in\R$, $(\id_{\{X\leq x\}},\W)\overset{d}{=}(\id_{\{X'\leq x\}},\W')$. It follows from
the weak law-invariance of $\{G_{\W}: \W\in\mathcal Y\}$ that $G_{\W}(1-\mathbb P(X\leq x|\W=\cdot))=G_{\W'}(1-\mathbb P(X'\leq x|\W'=\cdot))$. Hence by \eqref{eq:quantiles}, we have $\rho_{G_{\W}}(X,\W)=\rho_{G_{\W'}}(X',\W')$.

Next, we focus on the "only if" part. The LI of $\rho_{G_{\W}}(X,\W)$ implies that for $A, A'\in\mathcal F,\W, \W'\in\mathcal Y$ satisfying $(\id_A,\W)\overset{d}{=}(\id_{A'},\W')$, we have $\rho_{G_{\W}}(\id_A,\W)=\rho_{G_{\W'}}(\id_{A'},\W')$, which implies $G_{\W}(\mathbb P(A|\W=\cdot))=G_{\W'}(\mathbb P(A'|\W'=\cdot))=G_{\W'}(\mathbb P(A|\W=\cdot))$. Hence $\{G_{\W}: \W\in\mathcal Y\}$ is weakly law-invariant. \qed
\vspace{0.2cm}

{\bf Proof of Proposition \ref{prop:choquet}}. We first show the "if" part. By Theorem \ref{thm:choquet}, $\rho$ satisfies M, CA, N and LI. We next show that $\rho$ satisfies IC. For $A_{n}\uparrow A, n\to\infty$, we have $\mathbb{P}(A_n|\W)\uparrow \mathbb{P}(A|\W)$ a.s. as $n\to\infty$. We choose $f_n\in D^*, n\geq 1$  such that $f_n(\W)=\mathbb{P}(A_n|\W)$ a.s.. Let $g_n=\max_{i=1}^n f_i$ and $g=\sup_{i=1}^\infty f_i$. Then $g_n\uparrow g$ and $g_n(\W)=\mathbb{P}(A_n|\W)$ a.s. and $g(\W)=\mathbb{P}(A|\W)$ a.s..  Direct computation shows
\begin{align*}
1-F_{\id_{A_n}|\W=\w}(x)
=\left\{ \begin{array}{cc}
1, & x<0\\
\mathbb{P}(A_n|\W=\w), & 0\leq x<1\\
0, & x\geq1
\end{array}\right..
\end{align*}
By \eqref{eq:quantiles}, we have $\rho(\id_{A_n},\W)=\int_{0}^{1} G_{\W}\left(g_n\right) \d x=G_{\W}(g_n)$ and $\rho(\id_{A},\W)=G_{\W}(g)$. Using the fact that $g_n\uparrow g$ and the continuity from below of $G_{\W}$, we have $\lim_{n\to\infty}\rho(\id_{A_n},\W)=\lim_{n\to\infty} G_{\W}(g_n)=G_{\W}(g)=\rho(\id_A,\W)$, implying IC of $\rho$.

Next,  we focus on the "only if" part. By Theorem \ref{thm:choquet}, we have \eqref{eq:quantiles} holds for a law-invariant family of monotone functionals $\{G_{\W}: \W\in\mathcal Y\}$.
It suffices to show the continuity from below for $G_{\W}$. For $f_n\in D^*$, suppose  $f_n\uparrow f$.  Note that there exist $\W'$ and $U$ such that $\W'\overset{d}{=}\W$
and $U\sim U[0,1]$ is independent of $\W'$. Using the law-invariance of $\{G_{\W}: \W\in\mathcal Y\}$, we have $G_{\W'}(f_n)=G_{\W}(f_n)$ and $G_{\W'}(f)=G_{\W}(f)$. Hence, we only need to show $\lim_{n\to\infty}G_{\W'}(f_n)=G_{\W'}(f)$. Let $A_n'=\{U\leq f_n(\W')\}$. Then we have $\mathbb{P}(A_n'|\W')=f_n(\W')$ a.s.. The fact that $f_n\uparrow f$ implies $\mathbb{P}(\cup_{n=1}^\infty A_n'|\W')=f(\W')$ a.s.. It follows from \eqref{eq:quantiles} that $G_{\W'}(f_n)=\rho(\id_{A_n'},\W')$ and  $G_{\W'}(f)=\rho(\id_{\cup_{n=1}^\infty A_n'},\W')$. Hence, by M and IC of $\rho$ and the fact $A_n'\uparrow \cup_{n=1}^\infty A_n'$, we have $\lim_{n\to\infty }G_{\W'}(f_n)=\lim_{n\to\infty}\rho(\id_{A_n'},\W')=\rho(\id_{\cup_{n=1}^\infty A_n'},\W')=G_{\W'}(f)$.
\qed

\vspace{0.2cm}
{\bf Proof of Corollary \ref{prop:discrete}}.  Note that the proof of the "if" part is similar to the proof of Theorem \ref{thm:choquet}. Hence, it is omitted. We next show the "only if" part. By Theorem \ref{thm:choquet},
 there exists a law-invariant family of monotone functionals $\{G_{\W}:\W\in\mathcal{Y}\}$ such that
\begin{equation*}
\rho(X,\W)=\int_{0}^{\infty}G_{\W}\left(1-F_{X|\W=\cdot}(x)\right)\d x+\int_{-\infty}^{0}\left(G_{\W}\left(1-F_{X|\W=\cdot}(x)\right)-1\right)\d x.
\end{equation*}
For $\mathbf x\in [0,1]^n$, define $\psi_{\W}(\mathbf x)=G_{\W}(f)$, where $f(\w_i)=x_i$ for $i=1,\dots, n$, and $\W$ takes values $\w_1,\dots,\w_n$. If $\W'\overset{d}{=}\W$, then the values $\w_1,\dots,\w_n$ will be arranged in the same order.  Clearly, $\{\psi_{\W}:\W\in\mathcal Y\}$ is a law-invariant family of monotone functions. Moreover, by the definition, $\psi_{\W}\left(1-F_{X|\W=\w_1}(x),\dots, 1-F_{X|\W=\w_n}(x)\right)=G_{\W}(1-F_{X|\W=\cdot}(x))$ for all $x\in\R$. This implies the desired expression in Corollary \ref{prop:discrete}. \qed

\vspace{0.2cm}
\textbf{Proof of Proposition \ref{prop:coherent}}. By Theorem \ref{thm:choquet},
a mapping $\rho$ defined by \eqref{eq:quantiles} with a law-invariant family of monotone functionals $\{G_{\W}:\W\in\mathcal{Y}\}$
 satisfies M, CA and
N. Hence, in light of Theorem 4.94 in Follmer and Schied (2015), $\rho$
is coherent if and only if $\rho(\id_{A\cup B},\W)+\rho(\id_{A\cap B},\W)\leq\rho(\id_{A},\W)+\rho(\id_{B},\W)$
holds for all $A,B\in\mathcal{F}$ and $\W\in\mathcal{Y}$. We first focus on the "if" part.
 We could find $g_{1},g_{2},f_{1},f_{2}\in D^{*}$ such that $\mathbb{P}(A\cap B|\W)=g_{1}(\W)$,
$\mathbb{P}(A\cup B|\W)=g_{2}(\W)$, $\mathbb{P}(A|\W)=f_{1}(\W)$
and $\mathbb{P}(B|\W)=f_{2}(\W)$ hold almost surely. One can easily
check that $g_{1}(\W)\leq f_{1}(\W),f_{2}(\W)\leq g_{2}(\W)$ and
$f_{1}(\W)+f_{2}(\W)=g_{1}(\W)+g_{2}(\W)$ almost surely. We can choose
a version of $f_{1},f_{2},g_{1},g_{2}\in D^{*}$ such that $g_{1}\leq f_{1},f_{2}\leq g_{2}$
and $f_{1}+f_{2}=g_{1}+g_{2}$. Hence, by \eqref{eq:quantiles} and
Condition A, we have
\begin{align*}
\rho(\id_{A\cup B},\W)+\rho(\id_{A\cap B},\W) & =G_{\W}(\mathbb{P}(A\cup B|\W=\cdot))+G_{\W}(\mathbb{P}(A\cap B|\W=\cdot))\\
 & =G_{\W}(g_{1})+G_{\W}(g_{2})\leq G_{\W}(f_{1})+G_{\W}(f_{2})\\
 & =G_{\W}(\mathbb{P}(A|\W=\cdot))+G_{\W}(\mathbb{P}(B|\W=\cdot))\\
 & =\rho(\id_{A},\W)+\rho(\id_{B},\W).
\end{align*}
This implies $\rho$ is a coherent risk measure.

 We next focus on the "only
if" part. As $\rho$ is coherent, we have $\rho(\id_{A\cup B},\W)+\rho(\id_{A\cap B},\W)\leq\rho(\id_{A},\W)+\rho(\id_{B},\W)$
holds for all $A,B\in\mathcal{F}$ and $\W\in\mathcal{Y}$. It follows
from \eqref{eq:quantiles} that
\begin{align}\label{eq:11}
G_{\W}(\mathbb{P}(A\cup B|\W=\cdot))+G_{\W}(\mathbb{P}(A\cap B|\W=\cdot)) & \leq G_{\W}(\mathbb{P}(A|\W=\cdot))\nonumber \\
 & ~+G_{\W}(\mathbb{P}(B|\W=\cdot))
\end{align}
holds for all $A,B\in\mathcal{F}$ and $\W\in\mathcal{Y}$. We choose
a $\W\in\mathcal{Y}$ such that there exists $U\sim U[0,1]$ that
is independent of $\W$. For $f_{1},f_{2},g_{1},g_{2}\in D^{*}$ such
that $g_{1}\leq f_{1},f_{2}\leq g_{2}$ and $f_{1}+f_{2}=g_{1}+g_{2}$,
we set $B_{1}=\{U\leq g_{1}(\W)\}$, $A_{1}=\{U\leq f_{1}(\W)\}$,
$A_{2}=B_{1}\cup\{f_{1}(\W)<U\leq g_{2}(\W)\}$ and $B_{2}=\{U\leq g_{2}(\W)\}$.
It follows that $\mathbb{P}(A_{i}|\W)=f_{i}(\W)$ and $\mathbb{P}(B_{i}|\W)=g_{i}(\W)$
almost surely for $i=1,2$ and $B_{1}=A_{1}\cap A_{2}$ and $B_{2}=A_{1}\cup A_{2}$.
By \eqref{eq:11}, we have $G_{\W}(g_{1})+G_{\W}(g_{2})\leq G_{\W}(f_{1})+G_{\W}(f_{2})$.
In light of the law-invariance of $\{G_{\W}:\W\in\mathcal{Y}\}$, this
conclusion holds for all $\W\in\mathcal{Y}$. \qed

\subsection{Proof of results in Section \ref{sec:VaR}}\label{Appendix:VaR}

 \textbf{Proof of Theorem \ref{thm:MVaR}}. (i) We first show the
"if part". By \eqref{eq:3}, we have that M follows from the monotonicity of $D_{\W}$. For any strictly increasing and continuous function $\phi$, we
have
\begin{align*}
\rho\left(\phi(X),\W\right) & =\inf\{x:F_{\phi(X)|\W=\cdot}(x)\in_{\mathbb{P}\circ\W^{-1}}D_{\W}\}\\
 & =\inf\{x:F_{X|\W=\cdot}(\phi_{R}^{-1}(x))\in_{\mathbb{P}\circ\W^{-1}}D_{\W}\}\\
 & =\inf\{\phi(x):F_{X|\W=\cdot}(x)\in_{\mathbb{P}\circ\W^{-1}}D_{\W}\}\\
 & =\phi\left(\rho\left(X,\W\right)\right),
\end{align*}
where $\phi_{R}^{-1}(x)=\inf\{y:\phi(y)>x\}$ with the convention that $\inf\emptyset=\infty$.
Hence OR is satisfied. Moreover, the LI of $\rho$ is implied by the expression \eqref{eq:3} and the law-invariance of $\{D_{\W}:\W\in\mathcal Y\}$.

We next show the "only if" part. In light of Corollary 1 of \cite{C07}, and by M and OR, we have
\begin{align}
\rho\left(X,\W\right)=\int_{0}^{\infty}c^{\W}(X>x)\d x+\int_{-\infty}^{0}(c^{\W}(X>x)-1)\d x,\label{Eq:1}
\end{align}
where $c^{\W}$ is a capacity taking values from $\{0,1\}$. By Proposition \ref{le1}, we have $\rho$ satisfies CA.  Hence it follows from
Theorem \ref{thm:choquet} that there exists a law-invariant family of monotone
 functionals $\{G_{\W}:\W\in\mathcal{Y}\}$
such that \eqref{eq:quantiles} holds. For any strictly increasing and continuous
function $\phi$ with $\phi(0)=0$ and $\phi(1)=1$, we have $\phi(\id_{A})=\id_{A}$.
Hence we have $\rho(\id_{A},\W)=\phi(\rho(\id_{A},\W))$, implying
$\rho(\id_{A},\W)=0$ or $1$ for all $A\in\mathcal{F}$. Using \eqref{eq:quantiles},
we have $G_{\W}(\mathbb{P}(A|\W=\cdot))=\rho(\id_{A},\W)=0$ or $1$.
We let
\[
D_{\W}^{0}=\{f\in D^{*}:f\geq_{\mathbb{P}\circ\W^{-1}}\mathbb{P}(A|\W=\cdot)~\text{for some}~A\in\mathcal{F}~\text{such that}~\rho(\id_{A},\W)=1\}.
\]
By monotonicity of $G_{\W}$, we have $G_{\W}(f)=1$ for all $f\in D_{\W}^{0}$.
It follows from \eqref{eq:quantiles} that
\begin{align*}
\rho\left(X,\W\right) & =\int_{0}^{\infty}G_{\W}(1-F_{X|\W=\cdot}(x))\d x+\int_{-\infty}^{0}\left\{ G_{\W}(1-F_{X|\W=\cdot}(x))-1\right\} \d x\\
 & =\inf\{x\in\mathbb{R}:G_{\W}(1-F_{X|\W=\cdot}(x))=0\}\\
 & =\inf\{x\in\mathbb{R}:1-F_{X|\W=\cdot}(x)\in_{\mathbb{P}\circ\W^{-1}}D^{*}\setminus D_{\W}^{0}\}\\
 & =\inf\{x\in\mathbb{R}:F_{X|\W=\cdot}(x)\in_{\mathbb{P}\circ\W^{-1}}D_{\W}\},
\end{align*}
where $D_{\W}=1-D^{*}\setminus D_{\W}^{0}.$ Note that $1\in D_{\W}^{0}$,
$0\notin D_{\W}^{0}$ and $g\in D_{\W}^{0}$ if $g\geq f$ for some
$f\in D_{\W}^{0}$. Hence $1\in D_{\W}$, $0\notin D_{\W}$ and $g\in D_{\W}$
if $g\geq f$ for some $f\in D_{\W}$, which implies $D_{\W}$ is
an increasing set. Next, we show $\{D_{\W}:\W\in\mathcal{Y}\}$ can be chosen to be law-invariant. 
We fix $\W'\in\mathcal{Y}$ such that there exists $U\sim U[0,1]$ that is independent of $\W'$. Let $\W\overset{d}{=}\W'$.
For any $X\in \X$, there exists $X'\in\X$ such that $(X,\W)\overset{d}{=}(X',\W')$. By LI of $\rho$ and the above conclusion, we have
\begin{align*}
\rho\left(X,\W\right) & =\rho(X',\W')=\inf\{x\in\mathbb{R}:F_{X'|\W'=\cdot}(x)\in_{\mathbb{P}\circ\W^{-1}}D_{\W'}\}\\
&=\inf\{x\in\mathbb{R}:F_{X|\W=\cdot}(x)\in_{\mathbb{P}\circ\W^{-1}}D_{\W'}\}.
\end{align*}
 Hence,  $D_{\W}$ can be chosen as $D_{\W}=D_{\W'}$.

(ii) The proof of "if" part is similar to that of case (i). Next,
we only show the "only if" part. Let $Q_{\boldsymbol{\alpha},\BB}$
be the conditional probability measure $\mathbb{P}(\cdot|\mathrm{VaR}_{\boldsymbol{\alpha}}(\W)\leq\W\leq\mathrm{VaR}_{\BB}(\W))$
for $(\boldsymbol{\alpha},\BB)\in S_{\W}$. Define $f_{A}:S_{\W}\to[0,1]$
by $f_{A}(\A,\BB)=Q_{\A,\BB}(A)$. In light of Corollary 1 of \cite{C07}, and by M and OR, we have \eqref{Eq:1} holds.
 We denote
the set of all $f:S_{\W}\to[0,1]$ by $D_{\W}^{*}$ and let
\[
D_{\W}^{0}=\{f\in D_{\W}^{*}:f\geq f_{A}~\text{for some}~A\in\mathcal{F}~\text{such that}~c^{\W}(A)=1\}.
\]
It follows from \eqref{Eq:1} that $c^{\W}(A)=\rho\left(\mathbb{I}_{A},\W\right)$
for any $A\in\mathcal{F}$. Define $G_{\W}:D_{\W}^{*}\to\{0,1\}$
such that $G_{\W}(f)=1$ for $f\in D_{\W}^{0}$ and otherwise $G_{\W}(f)=0$.
Note that $\rho\left(X,\W\right)$ is law-invariant. Hence for $A,B\in\mathcal{F}$,
if $f_{A}=f_{B}$, then $(\mathbb{I}_{A},\W)\overset{d}{=}(\mathbb{I}_{B},\W)$,
which implies $c^{\W}(A)=c^{\W}(B)$.

Next we show that $c^{\W}(A)=1$ if and only if $f_{A}\in D_{\W}^{0}$.
It is obvious that $c^{\W}(A)=1$ implies $f_{A}\in D_{\W}^{0}$.
For the converse direction, note that $f_{A}\in D_{\W}^{0}$ implies
there exists $B\in\mathcal{F}$ such that $c^{\W}(B)=1$ and $f_{A}\geq f_{B}$.
Hence we have $\mathbb{P}(A\cap\{\W\in(\mathbf{a},\mathbf{b}]\})\geq\mathbb{P}(B\cap\{\W\in(\mathbf{a},\mathbf{b}]\})$
for all $\mathbf{a}=(a_{1},\dots,a_{N})$ and $\mathbf{b}=(b_{1},\dots,b_{N})$
with $a_{i}\leq b_{i},~i=1,\dots,N$. Define two measures $\mu_{A}$
and $\mu_{B}$ on $(\R^{N},\mathcal{B}(\R^{N}))$ by $\mu_{A}(\B)=\mathbb{P}(A\cap\{\W\in\B\})$
and $\mu_{B}(\B)=\mathbb{P}(B\cap\{\W\in\B\})$. Note that $\mathcal{C}:=\{(\mathbf{a},\mathbf{b}],~\mathbf{a}\leq\mathbf{b}\}$
is a semiring and $\mu_{A}(\B)\geq\mu_{B}(\B)$ for $\B\in\mathcal{C}$.
It follows from Theorem 11.3 of \cite{B95} that
\[
\mu_{E}(\B)=\inf\left\{ \sum_{i=1}^{\infty}\mu_{E}(\mathbf{C}_{i}):\mathbf{C}_{i}\in\mathcal{C}~\text{and}~\B\subset\cup_{i=1}^{\infty}\mathbf{C}_{i}\right\}
\]
holds for $E=A$ or $B$. Hence in light of the fact $\mu_{A}(\B)\geq\mu_{B}(\B)$
for all $\B\in\mathcal{C}$, we have $\mu_{A}(\B)\geq\mu_{B}(\B)$
for all $\B\in\mathcal{B}(\R^{N})$. This implies $\mathbb{P}(A|\W)\geq\mathbb{P}(B|\W)$
a.s.. Moreover, there exist $\W'$ and $U\sim U[0,1]$  such that $\W'\overset{d}{=}\W$
and $U$ is independent of $\W'$. Then $(F_{\mathbb{I}_{A}|\W'}^{-1}(U),\W')\overset{d}{=}(\mathbb{I}_{A},\W)$
and $(F_{\mathbb{I}_{B}|\W'}^{-1}(U),\W')\overset{d}{=}(\mathbb{I}_{B},\W)$,
where $F_{X|\boldsymbol{w}}^{-1}$ is the quantile function of $F_{X|\W=\boldsymbol{w}}$.
Note that $F_{\mathbb{I}_{A}|\W'}^{-1}(U)\geq F_{\mathbb{I}_{B}|\W'}^{-1}(U)$
a.s., and both of them are indicator functions. We let $A'=\{F_{\mathbb{I}_{A}|\W'}^{-1}(U)=1\}$
and $B'=\{F_{\mathbb{I}_{B}|\W'}^{-1}(U)=1\}$. Then we have $B'\subset A'$,
$(\id_{A'},\W')\overset{d}{=}(\id_{A},\W)$ and $(\id_{B'},\W')\overset{d}{=}(\id_{B},\W)$.
By LI, we have $c^{\W}(A)=\rho(\mathbb{I}_{A},\W)=\rho(\mathbb{I}_{A'},\W')$
and $c^{\W}(B)=\rho(\mathbb{I}_{B},\W)=\rho(\mathbb{I}_{B'},\W')$.
It follows from M and LI that $c^{\W}(A)=\rho(\mathbb{I}_{A'},\W')\geq\rho(\mathbb{I}_{B'},\W')=c^{\W}(B)=1.$
Consequently, $c^{\W}(A)=G_{\W}(f_{A})$ for $A\in\mathcal{F}$. Note
that $G_{\W}(f_{\{X>x\}})$ is decreasing with respect to $x$. It
follows from \eqref{Eq:1} that
\begin{align}
\rho\left(X,\W\right) & =\int_{0}^{\infty}G_{\W}(f_{\{X>x\}})\d x+\int_{-\infty}^{0}\left\{ G_{\W}(f_{\{X>x\}})-1\right\} \d x\nonumber \\
 & =\inf\{x\in\mathbb{R}:G_{\W}(f_{\{X>x\}})=0\}\nonumber \\
 & =\inf\{x\in\mathbb{R}:f_{\{X>x\}}\in D_{\W}^{*}\setminus D_{\W}^{0}\}=\inf\{x\in\mathbb{R}:f_{\{X\leq x\}}\in\widehat{D}_{\W}\}\nonumber \\
 & =\inf\left\{ x:\left(F_{X|\mathrm{VaR}_{\boldsymbol{\alpha}}(\W)\leq\W\leq\mathrm{VaR}_{\BB}(\W)}(x)\right)_{(\A,\BB)\in S_{\W}}\in\widehat{D}_{\W}\right\} ,
\end{align}
where $\widehat{D}_{\W}=1-D_{\W}^{*}\setminus D_{\W}^{0}.$ Note that
$1\in D_{\W}^{0}$, $0\notin D_{\W}^{0}$ and $g\in D_{\W}^{0}$ if
$g\geq f$ for some $f\in D_{\W}^{0}$. Hence $1\in\widehat{D}_{\W}$,
$0\notin\widehat{D}_{\W}$ and $g\in\widehat{D}_{\W}$ if $g\geq f$
for some $f\in\widehat{D}_{\W}$, which implies $\widehat{D}_{\W}$
is an increasing set. Similarly as in (i),  we can show that $\{\widehat{D}_{\W},~\W\in\mathcal{Y}\}$
can be chosen to be law-invariant.
\qed

\vspace{0.2cm}


\vspace{0.2cm}
{\bf Proof of Proposition \ref{VaR:weaklawinvariance}}. We first show the "if" part. If $(\mathbb{I}_{A},\W)\overset{d}{=}(\mathbb{I}_{A'},\W')$,
then we have $\mathbb P\circ \W^{-1}=\mathbb P\circ \W'^{-1}$ and
\begin{align}
F_{\mathbb{I}_{A}|\W=\w}(x)
=F_{\mathbb{I}_{A'}|\boldsymbol{W'}=\w}(x)=\left\{ \begin{array}{cc}
0, & x<0\\
\mathbb{P}(A^{c}|\W=\w), & 0\leq x<1\\
1, & x\geq1
\end{array}\right.\label{SC}
\end{align}
a.s. under $\mathbb P\circ \W^{-1}$. Direct computation
gives $\rho(\mathbb{I}_{A},\W)=\rho(\mathbb{I}_{A'},\W')=0$
if $\mathbb{P}(A^{c}|\W=\cdot)\in_{\mathbb P\circ \W^{-1}} D_{\W}\cap D_{\W'}$,
and $\rho(\mathbb{I}_{A},\W)=\rho(\mathbb{I}_{A'},\W')=1$
if $\mathbb{P}(A^{c}|\W=\cdot)\notin_{\mathbb P\circ \W^{-1}} D_{\W}\cup D_{\W'}$.
Hence, we have $\rho(\mathbb{I}_{A},\W)=\rho(\mathbb{I}_{A'},\W')$.
Moreover, by the proof of Theorem \ref{thm:MVaR}, $\rho$ satisfies CM and CA. Hence, in light of the second statement
of Proposition \ref{le1}, we obtain LI of $\rho$.

 Next, we consider the "only if" part. For $(\id_{A},\W)\overset{d}{=}(\id_{A'},\W')$, using \eqref{eq:3},  we have $\rho(\id_{A^c},\W)=\inf\{x\in\mathbb{R}: F_{\id_{A^c}|\W=\cdot}(x)\in_{\mathbb P\circ\W^{-1}} D_{\W}\}$ and $\rho(\id_{(A')^c},\W')=\inf\{x\in\mathbb{R}: F_{\id_{(A')^c}|\W'=\cdot}(x)\in_{\mathbb P\circ\W^{-1}} D_{\W'}\}$. Direct computation shows $\rho(\id_{A^c},\W)=0$ if $\mathbb P(A|\W=\cdot)\in_{\mathbb P\circ\W^{-1}} D_{\W}$ and otherwise $\rho(\id_{A^c},\W)=1$;  $\rho(\id_{(A')^c},\W')=0$ if $\mathbb P(A'|\W'=\cdot)\in_{\mathbb P\circ\W^{-1}} D_{\W'}$ and otherwise $\rho(\id_{A^c},\W')=1$.
Hence, the weak law-invariance of $\{D_{\W},~\W\in\mathcal{Y}\}$
is implied by the fact that $\rho(\mathbb{I}_{A^c},\W)=\rho(\mathbb{I}_{(A')^c},\W')$.  \qed

\subsection{Proof of results in Section \ref{sec:linear}.}

\textbf{Proof of Theorem \ref{thm:MES}}. We first show the "if"
part. First, we show \eqref{eq:MES} is well-defined. Let $f,g\in D^{*}$
such that $f(\W)=\mathbb{E}(X|\W)$ a.s. and $g(\W)=\mathbb{E}(X|\W)$
a.s., which imply $f=_{\mathbb{P}\circ\W^{-1}}g$. As $Q_{\W}<<\mathbb{P}\circ\W^{-1}$,
we have $f=_{Q_{\W}}g$. Consequently, $\mathbb{E}^{Q_{\W}}(f)=\mathbb{E}^{Q_{\W}}(g)$,
implying \eqref{eq:MES} is well-defined.

Next we show the properties of \eqref{eq:MES}. Clearly, $\rho$ satisfies M, AD and
N. For $(X,\W)\overset{d}{=}(X',\W')$, we have $\mathbb{E}(X|\W)\overset{d}{=}\mathbb{E}(X'|\W')$
and $\mathbb{P}\circ\W^{-1}=\mathbb{P}\circ\W'^{-1}$. Moreover, it
follows from the law-invariance of $\{Q_{\W}:\W\in\Y\}$
that $Q_{\W}=Q_{\W'}$. Note that $\E(X|\W=\cdot)=_{\mathbb{P}\circ\W^{-1}}\mathbb{E}(X'|\W'=\cdot)$
implies $\E(X|\W=\cdot)=_{Q_{\W}}\mathbb{E}(X'|\W'=\cdot)$.
Hence, $\mathbb{E}^{Q_{\W}}(\mathbb{E}(X|\W=\cdot))=\mathbb{E}^{Q_{\W'}}(\mathbb{E}(X'|\W'=\cdot))$,
implying the law-invariance of $\rho$. For $A_{n}\uparrow A$, we
have $\mathbb{E}(\id_{A_{n}}|\W)\uparrow\mathbb{E}(\id_{A}|\W)$ a.s..
It follows from the monotone convergence theorem that $\lim_{n\to\infty}\mathbb{E}^{Q_{\W}}(\mathbb{E}(\id_{A_{n}}|\W=\cdot))=\mathbb{E}^{Q_{\W}}(\mathbb{E}(\id_{A}|\W=\cdot))$.
Hence the IC is satisfied.

Next, we show the "only if" part. It follows from Proposition \ref{prop:choquet}
that \eqref{eq:quantiles} holds with a law-invariant family of monotone
and continuous from below functionals $\{G_{\W}:\W\in\mathcal{Y}\}$.
By \eqref{eq:quantiles}, we have $\rho(\id_{A},\W)=G_{\W}(\mathbb{P}(A|\W=\cdot))$
for all $A\in\mathcal{F}$. For any $\W\in\Y$, there exists $\W'\in\Y$
and $U\sim U[0,1]$ such that $\W'\overset{d}{=}\W$ and $\W'$ is
independent of $U$. By the law-invariance of $\{G_{\W}:\W\in\mathcal{Y}\}$,
we have $G_{\W'}=G_{\W}$. For $f,g\in D^{*}$ satisfying $f+g\in D^{*}$,
define $A=\{U\leq f(\W')\},B=\{f(\W')<U\leq f(\W')+g(\W')\}$ and
$C=A\cup B$. Then the additivity of $\rho$ implies $\rho(\id_{C},\W')=\rho(\id_{A},\W')+\rho(\id_{B},\W')$,
which is equivalent to $G_{\W'}(f+g)=G_{\W'}(f)+G_{\W'}(g)$. Hence,
we have $G_{\W}(f+g)=G_{\W}(f)+G_{\W}(g)$ holds for all $f,g\in D^{*}$
satisfying $f+g\in D^{*}$ and $\W\in\mathcal{Y}$. Moreover, by the
additivity and monotonicity of $G_{\W}$, we have $G_{\W}(af)=aG_{\W}(f)$
for $a\geq0,f\in D^{*}$ and $af\in D^{*}$.

Define $\Q_{\W}(\mathbf{B})=G_{\W}(\id_{\mathbf{B}})$ for all $\mathbf{B}\in\mathcal{B}(\R^{N})$.
The finite-additivity of $\Q_{\W}$ is implied by the additivity of
$G_{\W}$ and infinite-additivity of $\Q_{\W}$ is implied by the
continuity from below of $G_{\W}$. Hence, $\Q_{\W}$ is a probability
measure. Note that for $\mathbf{B}\in\mathcal{B}(\R^{N})$, if $\mathbb{P}(\W\in\mathbf{B})=0$,
then $\Q_{\W}(\mathbf{B})=G_{\W}(\id_{\mathbf{B}})=G_{\W}(0)=0$,
which implies $\Q_{\W}<<\mathbb{P}\circ\W^{-1}$. The law-invariance
of $\{G_{\W}:\W\in\Y\}$ implies the law-invariance of $\{\Q_{\W}:\W\in\Y\}$.
Let $f_{n}=\floor{2^nf}/2^n,~n\geq1$. Then one can easily check
that $G_{\W}(f_{n})=\E^{\Q_{\W}}(f_{n})$. Letting $n\to\infty$,
we have $f_{n}\uparrow f$. It follows from the continuity from below
for $G_{\W}$ that $\lim_{n\to\infty}G_{\W}(f_{n})=G_{\W}(f)$. Moreover,
the monotone convergence theorem implies $\lim_{n\to\infty}\E^{\Q_{\W}}(f_{n})=\E^{\Q_{\W}}(f)$.
Consequently, we have $G_{\W}(f)=\E^{\Q_{\W}}(f)$ for all $f\in D^{*}$.

Next, we show \eqref{eq:MES} holds. Note that $\rho(\id_{A},\W)=G_{\W}(\mathbb{P}(A|\W=\cdot))=\mathbb{E}^{Q_{\W}}(\mathbb{E}(\id_{A}|\W=\cdot))$
as $\mathbb{E}(\id_{A}|\W)=\mathbb{P}(A|\W)$ a.s.. Using the additivity
and monotonicity of both $\rho$ and $\mathbb{E}^{Q_{\W}}(\mathbb{E}(X|\W=\cdot))$,
we can show that \eqref{eq:MES} holds for $X_n=\floor{2^nX}/2^n,~n\geq1$. Note that $0\leq X-X_n\leq 2^{-n}$. Hence, $\lim_{n\to\infty}\rho(X_n,\W)=\rho(X,\W)$ and $\lim_{n\to\infty}\mathbb{E}^{Q_{\W}}(\mathbb{E}(X_n|\W=\cdot))=\mathbb{E}^{Q_{\W}}(\mathbb{E}(X|\W=\cdot))$, leading to
 the conclusion that \eqref{eq:MES} holds for all $X\in\X$. \qed
\subsection{Proof of results in \ref{sec:Coherent}}
{\bf Proof of Lemma \ref{le11}}. In light of  Lemma 3.3 of \cite{de24}, we have
\begin{align*}
{\mathrm{ess}\sup}_{Z\in L(X,\W)}\mathbb{E}\left(ZY|\W\right)=\int_{0}^{1}\VaR_{t}(X|\W)\VaR_t\left(Y|\W\right)\d t,
\end{align*}
where ${\mathrm{ess}\sup}$ is defined e.g., in Definition A.38 of \cite{FS16}.  This implies
$$\sup_{Z\in L(X,\W)}\mathbb{E}\left(ZY\right)=\sup_{Z\in L(X,\W)}\mathbb{E}\left(\mathbb{E}\left(ZY|\W\right)\right)\leq\mathbb{E}\left(\int_{0}^{1}\VaR_{t}(X|\W)\VaR_t\left(Y|\W\right)\d t\right).$$
We next show the inverse inequality by following the same idea as in the proof of Lemma 3.3 of de Castro et al.(2024).  If $F_{Y|\W}(x)$ is a continuous function with respect to $x\in\R$ on $\Omega_1\in \mathcal F$ with $\mathbb P(\Omega_1)=1$, then for $t\in (0,1)$, $\mathbb P(F_{Y|\W}(Y)\geq t|\W)=\mathbb P(Y\geq \VaR_t(Y|\W)|\W)=1-t$ a.s.. Hence, $\mathbb P(F_{Y|\W}(Y)\in\cdot|\W)\sim U(0,1)$ a.s..  Let $V=F_{Y|\W}(Y)$. It follows that $\mathbb P(\VaR_{V}(X|\W)\leq x|\W)=\mathbb P(V\leq F_{X|\W}(x)|\W)=F_{X|\W}(x)$ a.s. for all $x\in\R$. Hence, $(\VaR_{V}(X|\W),\W)\overset{d}{=}(X,\W)$ implying $\VaR_{V}(X|\W)\in L(X,\W)$. By definition,  $Y\geq \VaR_{V}(Y|\W)$ a.s.. Moreover, direct computation shows $$\mathbb E(Y)=\mathbb{E}(\mathbb E(Y|\W))=\mathbb E\left(\int_{0}^{1}\VaR_t(Y|\W)\d t\right)=\mathbb E(\VaR_{V}(Y|\W)).$$ Hence, we have $Y=\VaR_{V}(Y|\W)$ a.s.. Using the above conclusion, we obtain $$\mathbb{E}\left(\VaR_{V}(X|\W)Y|\W\right)=\mathbb{E}\left(\VaR_{V}(X|\W)\VaR_{V}(Y|\W)|\W\right)=\int_{0}^{1}\VaR_{t}(X|\W)\VaR_t\left(Y|\W\right)\d t.$$
Hence, $$\sup_{Z\in L(X,\W)}\mathbb{E}\left(ZY\right)=\mathbb{E}\left(\int_{0}^{1}\VaR_{t}(X|\W)\VaR_t\left(Y|\W\right)\d t\right).$$
Without loss of generality, we suppose $X\geq 0$ in the following proof.
For $Y\in L^1$, there exists a sequence of discrete random variables $Y_n\in L^1$ such that $Y_n\geq Y$ and $|Y_n-Y|\leq 1/n$. Let $U\sim U(0,1)$ be independent of $\W$. It follows that $\mathbb P(U\in\cdot|\W)\sim U(0,1)$ a.s.. One can easily check that $\mathbb P(Y_n+U/n\leq x|\W)$ is continuous over $\R$ on some $\Omega_2\in\mathcal F$ with $\mathbb P(\Omega_2)=1$. Using the above conclusion, there exists $X_n\in L(X,\W)$ such that
$$\mathbb{E}\left(X_n(Y_n+U/n)|\W\right)=\int_{0}^{1}\VaR_{t}(X|\W)\VaR_t\left(Y_n+U/n|\W\right)\d t~ a.s..$$
Consequently,
\begin{align*}\mathbb{E}\left(\int_{0}^{1}\VaR_{t}(X|\W)\VaR_t\left(Y|\W\right)\d t\right)&\leq \mathbb{E}\left(\int_{0}^{1}\VaR_{t}(X|\W)\VaR_t\left(Y_n+U/n|\W\right)\d t\right)\\
&=\mathbb{E}\left(X_n(Y_n+U/n)\right)\\
&\leq \sup_{Z\in L(X,\W)}\mathbb{E}\left(ZY\right)+\mathbb{E}(X_n)/n+\mathbb{E}\left(X_nU\right)/n.
\end{align*}
Letting $n\to\infty$, we obtain $$\mathbb{E}\left(\int_{0}^{1}\VaR_{t}(X|\W)\VaR_t\left(Y|\W\right)\d t\right)
\leq \sup_{Z\in L(X,\W)}\mathbb{E}\left(ZY\right),$$
which is the desired inverse inequality.

Note that
\begin{align*}
\inf_{Z\in L(X,\W)}\mathbb{E}\left(ZY\right)=\inf_{Z\in L(-X,\W)}\mathbb{E}\left(-ZY\right)
=-\sup_{Z\in L(-X,\W)}\mathbb{E}\left(ZY\right).\end{align*}
Hence, using the above conclusion, we have
\begin{align*}
\inf_{Z\in L(X,\W)}\mathbb{E}\left(ZY\right)&=-\mathbb{E}\left(\int_{0}^{1}\VaR_{t}(-X|\W)\VaR_t\left(Y|\W\right)\d t\right)\\
&=\mathbb{E}\left(\int_{0}^{1}\VaR_{1-t}(X|\W)\VaR_{t}\left(Y|\W\right)\d t\right).
\end{align*}
\qed

\vspace{0.2cm}
{\bf Proof of Theorem \ref{prop:distortion-distribution-1}}.
It follows from Corollary 4.38 of \cite{FS16} that $\rho$ is coherent and continuous from above if and only if
\begin{align*}
\rho\left(X,\W\right) =\sup_{Q\in \mathcal{Q}_{\W}}\mathbb{E}\left(X\frac{\d Q}{\d \mathbb P}\right)
\end{align*}
for some $\mathcal{Q}_{\W}\subset \mu_1(\mathbb P)$. Next, we only need to show that $\rho$ is additionally law-invariant if and only if
\eqref{coherent} holds. The "if" part is obvious. In light of \eqref{coherent} and the law-invariance of $\{\mathcal Q_{\W}: \W\in \mathcal Y\}$, $(X,\W)\overset{d}{=}(X',\W')$ implies $\rho(X,\W)=\rho(X',\W')$. Hence, $\rho$ is law-invariant.  We next show the "only if" part.
Using the law-invariance of $\rho$, we have
\begin{align*}
\rho\left(X,\W\right) =\sup_{Z\in L(X,\W)}\rho(Z,\W)=\sup_{Q\in \mathcal{Q}_{\W}}\sup_{Z\in L(X,\W)}\mathbb{E}\left(Z\frac{\d Q}{\d \mathbb P}\right).
\end{align*}
By Lemma \ref{le11}, we have
\begin{align*}
\rho\left(X,\W\right)=\sup_{Q\in \mathcal{Q}_{\W}}\mathbb{E}\left(\int_{0}^{1}\VaR_{t}(X|\W)\VaR_t\left(\frac{\d Q}{\d \mathbb P}\Big|\W\right)\d t\right).
\end{align*}
Suppose $\W'\in \mathcal Y$ such that $\W'\overset{d}{=}\W$. Then for $X'\in\X$, there exists $X\in\X$ such that $(X',\W')\overset{d}{=}(X,\W)$. Hence, \begin{align*}
\rho(X',\W')=\rho(X,\W)&=\sup_{Q\in \mathcal{Q}_{\W}}\mathbb{E}\left(\int_{0}^{1}\VaR_{t}(X|\W)\VaR_t\left(\frac{\d Q}{\d \mathbb P}\Big|\W\right)\d t\right)\\
&=\sup_{Q\in \mathcal{Q}_{\W'}}\mathbb{E}\left(\int_{0}^{1}\VaR_{t}(X'|\W')\VaR_t\left(\frac{\d Q}{\d \mathbb P}\Big|\W'\right)\d t\right)
\end{align*}
if $\{F_{Z, \W'}: Z=\frac{\d Q}{\d \mathbb{P}},~Q\in \mathcal{Q}_{\W'}\}=\{F_{Z, \W}: Z=\frac{\d Q}{\d \mathbb{P}},~Q\in \mathcal{Q}_{\W}\}$. Hence, $\{\mathcal{Q}_{\W}:\W\in\mathcal Y\}$ can be chosen to be law-invariant.
 \qed
 \subsection{Proof of results in Section \ref{sec:risksharing}}

 {\bf Proof of Proposition \ref{prop:risksharing}}. Fix $U\sim U[0,1]$ such that $U$ and $X$ are comonotonic. Let $\Lambda_{i}(1-x)=G_{\W_i}(\mathbb P(U>x|\W_i=\cdot))$ for $x\in [0,1]$. Clearly, $\Lambda_i:[0,1]\to [0,1]$ is an increasing function with $\Lambda_i(0)=0$ and $\Lambda_i(1)=1$. Since $G_{\W_i}$ is continuous from below, then $\Lambda_i$ is left-continuous. For $(X_1,\dots, X_n)\in \mathbb A_n^+(X)$, there exist increasing and Lipschitz continuous functions $h_1,\dots,h_n$ such that $X_1=h_1(X),\dots, X_n=h_n(X)$ and $h_1(x)+\cdots+h_n(x)=x,~x\in\R$. Direct computation gives
$G_{\W_i}(\mathbb P(h_i(X)>x|\W_i=\cdot))=\Lambda_i(1-F_{h_i(X)}(x))$ for all $x\in\R$. Hence, we have
\begin{align*}
\sum_{i=1}^n\rho_{G_{\W_i}}(X_i,\W_i)&=\sum_{i=1}^{n}\left(\int_{0}^{\infty}\Lambda_i(1-F_{h_i(X)}(x))\d x+\int_{-\infty}^{0}\left(\Lambda_i(1-F_{h_i(X)}(x))-1\right)\d x\right)\\
&=\sum_{i=1}^n\left(\int_{0}^{\infty}\Lambda_i(1-F_{X}(x))h_i'(x)\d x+\int_{-\infty}^{0}\left(\Lambda_i(1-F_{X}(x))-1\right)h_i'(x)\d x\right)\\
&\geq \int_{0}^{\infty} G_{\W_1,\dots,\W_n,X}(x)\d x+\int_{-\infty}^{0}\left(G_{\W_1,\dots,\W_n,X}(x)-1\right)\d x.
\end{align*}
The above inequality becomes equality if $h_i'(x)=0$ when $G_{\W_i}(1-F_{X|\W_i=\cdot}(x))>G_{\W_1,\dots,\W_n,X}(x)$, and $\sum_{i=1}^{n} h_i'(x)=1,~x\in\R$. We establish the claim.  \qed
\subsection{A counterexample}\label{cexample}
\begin{example}\label{example}
Let $([0,1], \mathcal{B}([0,1]), \lambda)$ be the probability space, where $\mathcal{B}([0,1])$ is the set of all Borel subsets of $[0,1]$ and $\lambda$ is the Lebesgue measure. Moreover, let $\X=L^{\infty}$ and $\mathcal Y=\{U: U\sim U[0,1]\}$. Let $U_0(x)=x,~x\in [0,1]$.  For $U\in \mathcal Y\setminus\{U_0\}$, define  $G_{U}(f)=\id_{D_{U}}(f)$ for all $f\in D^*$, where $D_{U}=\{f\in D^*: \lambda(f(U)\geq 1/2)=1\}$.  Moreover, let $G_{U_0}(f)=\id_{D_{U_0}}(f)$ with $D_{U_0}=\{f\in D^*: \lambda(f(U_0)> 1/2)=1\}$. One can easily check that for $U\in\mathcal Y$, $G_U$ is monotone. For $(\id_A,U)\overset{d}{=}(\id_{A'},U')$, there exists $g\in D^*$ such that $g(U)=\lambda(A|U)$ and $g(U')=\lambda(A'|U')$ a.s.. Hence, $G_U(g)=G_{U'}(g)$ if $U, U'\in \mathcal Y\setminus\{U_0\}$ or $U=U'=U_0$. Now we consider the case $U\neq U_0$ and $U'=U_0$. If $\lambda(g(U')>1/2)=1$, then $\lambda(g(U)\geq 1/2)=1$, implying $G_{U}(g)=G_{U'}(g)=1$. Suppose $\lambda(g(U')>1/2)<1$ and $\lambda(g(U)\geq 1/2)=1$. Then $\lambda(g(U')=1/2)=\lambda(g(U)=1/2)>0$, which implies there exists $B\in \mathcal{B}([0,1])$ such that $\lambda(A'|U')=1/2$ over $\{U'\in B\}$ with $\lambda(B)>0$. It follows that $\lambda(A'\cap \{U'\in C\})=\frac{1}{2}\lambda(\{U'\in C\})$ for all $C\in \mathcal{B}([0,1])$  and $C\subset B$. Using the fact that $\{U'\in C\}=C$ for $C\in \mathcal{B}([0,1])$, we have $\lambda(A'\cap C)=\frac{1}{2}\lambda(C)$ for all $C\in \mathcal{B}([0,1])$ and $C\subset B$. Choosing $C=A'\cap B$, we have $\lambda(A'\cap B)=\frac{1}{2}\lambda(A'\cap B)$. Thus we have $\lambda(A'\cap B)=0$. However, for $C=B$, $\lambda(A'\cap B)=\frac{1}{2}\lambda(B)>0$, leading to a contradiction. Hence, if $\lambda(g(U')>1/2)<1$, then $\lambda(g(U)\geq 1/2)<1$, implying $G_{U}(g)=G_{U'}(g)=0$. Consequently, we conclude that $\{G_U: U\in\mathcal Y\}$ is weakly law-invariant. Note that $G_{U}(1/2)=1$ if $U\neq U_0$ and $G_{U_0}(1/2)=0$. Hence, $\{G_U: U\in\mathcal Y\}$ is not law-invariant.
\end{example}

\end{document}